\documentclass[useAMS,usenatbib]{mn2e}

\usepackage[T1]{fontenc}
\usepackage{ae,aecompl}
\usepackage{times} 

\usepackage{float}
\usepackage{graphicx}
\usepackage{dcolumn}
\usepackage{bm}
\usepackage{listings}
\usepackage{amssymb}
\usepackage{verbatim}
\usepackage{url}
\usepackage{amsmath}
\usepackage{longtable}
\usepackage{booktabs}
\usepackage{ctable}
\usepackage{natbib}
\usepackage{hyperref}

\makeatletter
\newcommand{\HI}{{\rm H\,{\scriptstyle I}}}
\newcommand{\HII}{{\rm H\,{\scriptstyle II}}}

\newcommand{\Rmnum}[1]{\expandafter\@slowromancap\romannumeral #1@}
\newcommand{\nH}{n_{\mbox{\tiny H}}}
\newcommand{\nHI}{n_{\mbox{\tiny H\Rmnum{1}}}}
\newcommand{\xHI}{x_{\mbox{\tiny H\Rmnum{1}}}}

\newcommand{\nHII}{n_{\mbox{\tiny H\Rmnum{2}}}}

\newcommand{\LyA}{\mbox{Ly}\alpha}

\newcommand{\NHI}{N_{\mbox{\tiny H\Rmnum{1}}}}
\newcommand{\fHI}{\langle f_{\mbox{\tiny H\Rmnum{1}}}\rangle}

\makeatother

\title[$\LyA$-emitting galaxies as a probe of reionization]
  {$\LyA$-Emitting Galaxies as a Probe of Reionization: \\ Large-Scale Bubble Morphology and Small-Scale Absorbers}  

\author[K. Kakiichi et al.]{Koki Kakiichi,$^{1}$\thanks{E-mail: kakiichi@mpa-garching.mpg.de}
Mark Dijkstra,$^{1,2}$
Benedetta Ciardi,$^{1}$
Luca Graziani$^{1,3}$
\\
$^1$Max Planck Institute for Astrophysics, Karl-Schwarzschild Stra\ss e 1, 85741 Garching, Germany
\\
$^2$Institute of Theoretical Astrophysics, University of Oslo, P.O box 1029, Blindern, 0315 Oslo, Norway 
\\
$^3$INAF Osservatorio Astronomico di Roma, Via Frascati 33, 00040, Monte Porzio Catone (RM), Italy
}

\date{Released 2015 Xxxxx XX}

\pagerange{\pageref{firstpage}--\pageref{lastpage}} \pubyear{2015}

\def\LaTeX{L\kern-.36em\raise.3ex\hbox{a}\kern-.15em
    T\kern-.1667em\lower.7ex\hbox{E}\kern-.125emX}

\begin{document}

\maketitle

\begin{abstract}
The visibility of $\LyA$-emitting galaxies during the Epoch of Reionization (EoR) is controlled by both diffuse $\HI$ patches in large-scale bubble morphology and small-scale absorbers. 
To investigate their impact on $\LyA$ photons, we apply a powerful and novel combination of analytic and numerical calculations to three scenarios: ({\it i}) the `bubble' model, where only diffuse $\HI$ outside ionized bubbles is present; ({\it ii}) the `web' model, where $\HI$ exists only in overdense self-shielded gas; and ({\it iii}) the more realistic 'web-bubble' model, which contains both. 
Our analysis confirms that there exists a degeneracy between the ionization structure of the intergalactic medium (IGM) and the global $\HI$ fraction inferred from $\LyA$ surveys, as the three models suppress $\LyA$ flux equally with (very) different neutral fractions.
We argue that a joint analysis of the $\LyA$ luminosity function and the rest-frame equivalent width distribution/$\LyA$ fraction of $\LyA$- and UV-selected galaxies can break this degeneracy and provide constraints on the reionization history and its topology.  
We further show that constraints can improve if we consider the full shape of the $M_{UV}$-dependent redshift evolution of the $\LyA$ fraction of Lyman break galaxies. Contrary to conventional wisdom, we find that ({\it i}) a drop of $\LyA$ fraction larger
for UV-faint than for UV-bright galaxies can be reproduced with web and web-bubble models and therefore does not provide exclusive evidence of patchy reionization, and ({\it ii}) the IGM-transmission PDF is unimodal for bubble models and bimodal in web models. Our analysis further highlights the importance of galaxy-absorber cross-correlation and its systematic measurement from the post-reionized era to the EoR. Comparing our grid of models to observations, the neutral fraction at $z\sim 7$ is likely to be of order of tens of per cent when interpreted with bubble or web-bubble models. Alternatively, we obtain a conservative lower limit on the neutral fraction of $\sim 1\%$ in the web models, if we allow for a drop in the photoionization rate by a factor of $\sim100$ from the post-reionized universe.
\end{abstract}


\section{Introduction}

The Epoch of Reionization (EoR) and Cosmic Dawn are the least explored frontiers in observational cosmology and extragalactic astrophysics (\citealt{2013fgu..book.....L}). Galaxy surveys are one of the most important pillars of modern cosmology, allowing us to study high-redshift galaxy formation and the reionization process of the intergalactic medium (IGM). Surveys of high-redshift galaxies using Lyman-break drop-out technique (Lyman Break Galaxies, LBGs) (e.g.  \citealt{2011MNRAS.418.2074M}; \citealt{2013ApJ...763L...7E}; \citealt{2015ApJ...803...34B}) and narrow-band filter targeting $\LyA$ emission (Lyman Alpha Emitters, LAEs) (e.g. \citealt{2004ApJ...617L...5M}; \citealt{2010ApJ...725..394H}; \citealt{2010ApJ...723..869O}) have provided a deep sample of objects, indicating that reionization requires many faint galaxies below the sensitivity limit of the surveys (\citealt{2013ApJ...768...71R}). Furthermore, observations of QSO spectra (e.g. \citealt{2006AJ....132..117F}; \citealt{2015MNRAS.447.3402B}) and the Cosmic Microwave Background (CMB) (\citealt{2013ApJS..208...19H}; \citealt{2015arXiv150201589P}; \citealt{2012ApJ...756...65Z}) offer hints that reionization is mostly completed at $z\gtrsim6$. 

However, beyond such indications, our present observational constraints on the EoR are still scarce, regarding both the reionization history and its topology/morphology. While 21cm experiments with radio interferometers such as 
LOFAR\footnote{\url{http://www.lofar.org}}, 
MWA\footnote{\url{http://www.mwatelescope.org}},
GMRT\footnote{\url{http://gmrt.ncra.tifr.res.in}}, 
PAPER\footnote{\url{http://astro.berkeley.edu/\~dbacker/eor}},
HERA\footnote{\url{http://reionization.org}},
SKA\footnote{\url{http://www.skatelescope.org}}
offer the most direct probe of the physical state of the IGM during the EoR (e.g. \citealt{2012RPPh...75h6901P}), the challenge in foreground removal and calibration remains. Substantial progress has been recently made by the 21cm community (e.g. \citealt{2013MNRAS.433..639P}; \citealt{2014ApJ...788..106P}; \citealt{2013A&A...550A.136Y}), but a detection is still missing. In principle, surveys of $\LyA$-emitting galaxies ($\LyA$ surveys) offer an alternative and independent means from 21cm experiments to probe the EoR and constrain the global $\HI$ fraction (e.g. \citealt{2014PASA...31...40D}). Such an approach is attractive because of the present availability of data and up-coming surveys with the Hyper Suprime-Cam on Subaru\footnote{\url{http://www.naoj.org/Projects/HSC/}} and with future telescopes such as JWST\footnote{\url{http://www.jwst.nasa.gov/}}, E-ELT\footnote{\url{http://www.eso.org/public/teles-instr/e-elt/}}, TMT\footnote{\url{http://www.tmt.org/}}, and GMT\footnote{\url{http://www.gmto.org/}}.  Furthermore, using multiple independent strategies can provide constraints on reionization which are less sensitive to systematic uncertainties of individual probes.

The challenge in using Ly$\alpha$ emitting galaxies as a probe of reionization lies in correctly interpreting observations (\citealt{2012ApJ...744...83O}; \citealt{2012MNRAS.422.1425C}; \citealt{2010MNRAS.408.1628S}; \citealt{2013ApJ...775L..29T}; \citealt{2014MNRAS.443.2831C}; \citealt{2014ApJ...794....5T}; \citealt{2015A&A...573A..24C}). The reduced visibility of $\LyA$ emission from galaxies at $z>6$ has already been used to infer the global $\HI$ fraction of the IGM (e.g. \citealt{2011MNRAS.414.2139D}; \citealt{2014MNRAS.444.2114J}). However, a robust interpretation is still uncertain because of the complex radiative transfer of both ionizing and Ly$\alpha$ photons. The Ly$\alpha$ transfer involves a wide range of scales including ({\it i}) the interstellar medium (ISM), where dust and gas distribution and kinematics determine the escape fraction of Ly$\alpha$ photons as well as their spectral line profile (e.g \citealt{2006A&A...460..397V}; \citealt{Max}; \citealt{2014MNRAS.441.2861H}); ({\it ii}) the circum-galactic medium (CGM), i.e. the direct environment of galaxies out to a few hundred kpc \citep[e.g.][]{DIGM,2010ApJ...716..574Z, 2011ApJ...726...38Z,2011ApJ...728...52L}; and ({\it iii}) the IGM, which - during reionization - contains diffuse neutral gas surrounding large ionized bubbles which themselves contain dense, self-shielding gas clouds. In order to obtain robust constraints on the global $\HI$ fraction, it is essential to understand the cosmological $\LyA$ RT on all these scales.

The precise ionization structure of the IGM, i.e. the topology of reionization, is not only characterized by the size, abundance and distribution of large-scale ionized bubbles, but also by the small-scale dense $\HI$ absorbers self-shielded against the external ionizing sources. Interpretations of $\LyA$-emitting galaxies contain (often implicit) assumptions about the ionization structure of the IGM, mostly because of the difficulty to cover the entire dynamic range that is required to properly describe both the small-scale dense $\HI$ absorbers \textit{and} the large-scale diffuse neutral IGM in reionization simulations. Two extreme assumptions, described in the first two bullets below, have been commonly adopted in the literature. Here we introduce the following terminology:
\begin{itemize}
\item {\it Bubble model}: 
in this model small-scale $\HI$ absorbers are neglected. Under this assumption, the global $\HI$ fraction measures the $\HI$ content of the diffuse neutral IGM outside ionized bubbles.We refer to this as the `bubble model'.
\item {\it Web model}: 
here only the small-scale $\HI$ absorbers are considered. As this overdense gas largely traces the large-scale cosmic web, we refer to it as the `web model'.
\item {\it Web-bubble model}: 
reality is a combination of the two extreme configurations above. We refer to cases that contain both neutral phases (diffuse and clumped) of gas as the hybrid `web-bubble model'. One can visualize this as the more common bubble model, but with `impurities' in the ionized bubbles in the form of small-scale neutral islands. 
\end{itemize}

\begin{figure*}
  \includegraphics[angle=-90,width=\textwidth]{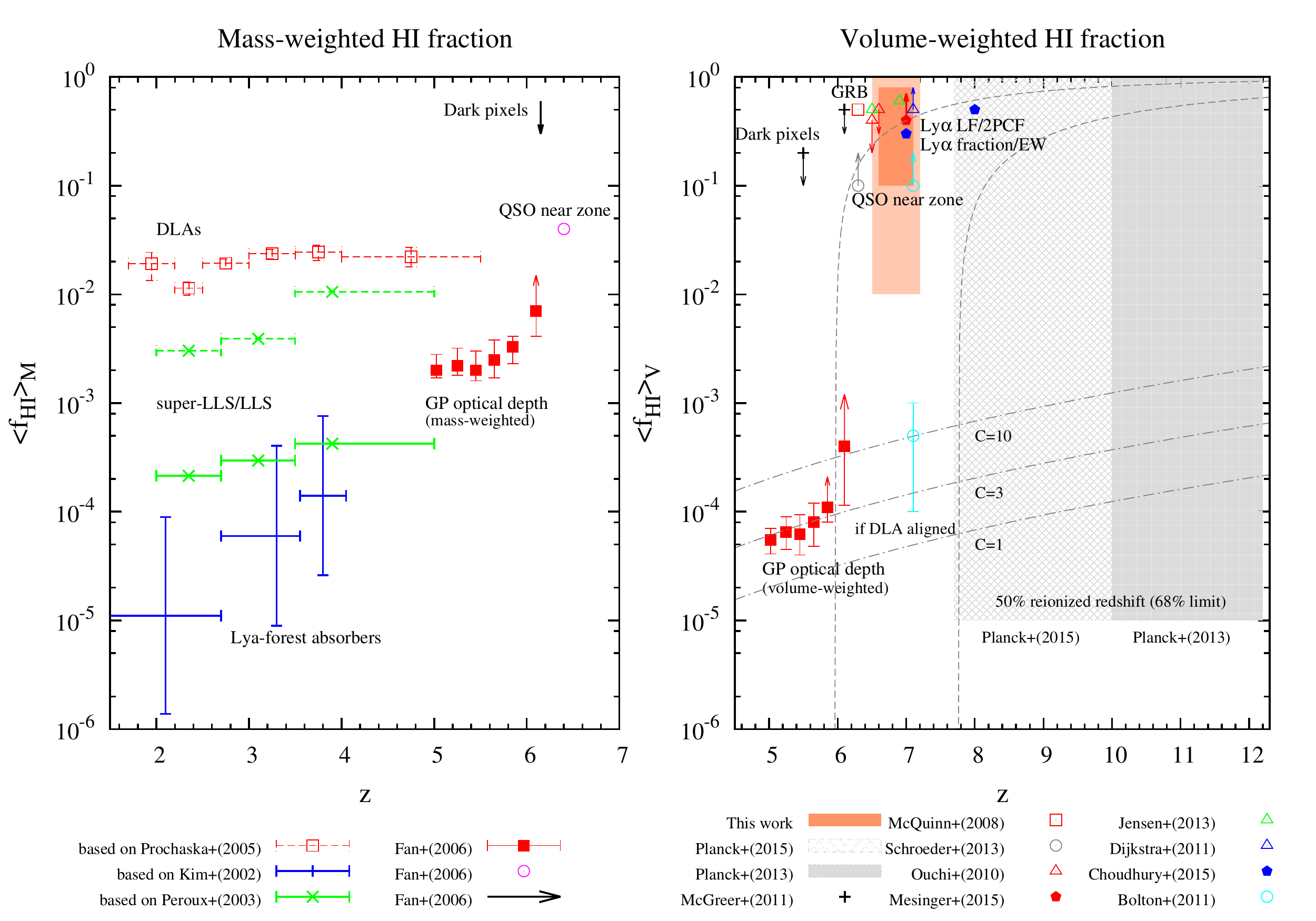}
  \caption{Cosmological $\HI$ fraction $\fHI_{V,M}$ in the diffuse IGM and high-column density $\LyA$ absorbers (LLS/DLAs) at various redshifts, from the post-reionized universe to the epoch of reionization. The mass-weighted $\HI$ fraction embedded in the small-scale absorbers is computed from equation (\ref{HIfromCDDF}) using the fitting function to the observations of Prochaska et al. (2005), Kim et al. (2002) and Peroux et al. (2003). All the other values are compiled from the literature, as indicated by the labels above (z=7 constraints by Dijkstra et al. (2011) and Jensen et al. (2013) are shifted by +0.1 and -0.1 in redshift to avoid a cluttering of data points). The dash-dotted lines in the right panel are the volume-weighted neutral fraction of a diffuse IGM with clumping factor $C=1,3,10$ in photoionization equilibrium with a UV background $\Gamma=10^{-12}$~s$^{-1}$. The dashed lines are the bubble model estimates of the neutral fraction in $\HI$ patches outside ionized bubbles (left and right lines are DEC and CONST models in \S~\ref{sec:bubble}).
}\label{fig:HIcotent}
\end{figure*}

Most previous works interpreting the observed reduction in $\LyA$ flux from $z>6$ galaxies have favoured a very high value of $\HI$ fraction, as high as $\sim50\%$ at $z\sim7$ (e.g. \citealt{2007MNRAS.381...75M}; \citealt{2011MNRAS.414.2139D}; \citealt{2013MNRAS.428.1366J}). These studies used large-scale reionization simulations which did not have the spatial resolution to resolve the self-shielded small-scale absorbers.

The lack of self-shielding gas inside ionized bubbles in large-scale reionization simulations is clearly problematic: $\LyA$ forest observations indicate that in the post-reionized universe, i.e. $z<5$, $\HI$ gas is locked up in damped $\LyA$ systems (DLA) and Lyman-limit systems (LLS) (e.g. \citealt*{2005ARA&A..43..861W}). Self-shielded absorbers (LLSs and DLAs) are also expected to reside inside ionized bubbles during reionization (and possibly with larger number densities, see e.g. \citealt{2013MNRAS.429.1695B}). The first investigations of hybrid web-bubble models have recently been reported (\citealt{2015MNRAS.446..566M}; \citealt{2015MNRAS.452..261C}). Interestingly, these papers still favour large values for the $\HI$ fraction, as high as $\sim40\%$ at redshift $z=7$.

In this paper we investigate the impact of large-scale patchy reionization and small-scale $\HI$ absorbers on the observed $\LyA$ flux of galaxies, and its implication on the $\HI$ fraction measurements from $\LyA$ surveys.  We explore a unique combination of cosmological hydrodynamical, radiative transfer simulations and analytic models. Our analytic framework is powerful as it facilitates the interpretation of the results of our simulations, and provides us with a tool to quickly explore a large range of parameters describing the reionization and $\LyA$ transfer processes in future work.

The paper is organized as follows. Section \ref{sec:HIcontent} briefly reviews the cosmic history of the $\HI$ content in the universe, ranging from the epoch of reionization to the post-reionized universe. In Section \ref{analytic} we present our analytic framework of cosmological $\LyA$ radiative transfer. In Section \ref{sec:sims} we describe the methodology employed to generate the reionization models (bubble, web, and web-bubble models), as well as the intrinsic and apparent mock catalogue of $\LyA$-emitting galaxies. Section \ref{sec:result} shows our results. The conclusions and discussion about implications on $\LyA$-emitting galaxy surveys are then presented in Section \ref{sec:conclusion}.

\section{Cosmological $\HI$ content}\label{sec:HIcontent}

In this section, we review the redshift evolution of the $\HI$ content both during and after reionization. This can be quantified either by the mass-weighted $\fHI_{M}$ or the volume-weighted $\fHI_{V}$ neutral fraction. A compilation of current estimates in the literature is shown in Fig.~\ref{fig:HIcotent}. 

\subsection{Observational Constraints on $\HI$ in the Post-Reionization Epoch}
The left panel of Fig.~$\ref{fig:HIcotent}$ clearly indicates that the post-reionized universe still contains neutral islands of gas in the form of self-shielding LLSs and DLAs. The abundance of the $\HI$ gas is generally quantified in terms of the $\HI$ column density distribution function (CDDF), $f(\NHI,z)$, which is defined as (e.g. \citealt*{2005ARA&A..43..861W}) $f(\NHI,z)=\frac{\partial^2 \mathcal{N}}{\partial\NHI\partial z}\frac{H(z)}{H_0(1+z)^2},$ where $\frac{\partial^2 \mathcal{N}}{\partial\NHI\partial z}$ is  the number of $\LyA$ absorbers $\mathcal{N}(\NHI,z)$ per unit $\HI$ column density $\NHI$ and per unit redshift, $H(z)=H_0[\Omega_m(1+z)^3+\Omega_\Lambda]^{1/2}$ and $H_0$ is the Hubble parameter today. The mass-weighted $\HI$ fraction embedded in small-scale absorbers is estimated from observations of $f(\NHI,z)$ (see Appendix \ref{A1})\footnote{Converting $f(\NHI,z)$ into a constraint on the volume-weighted $\HI$ fraction, $\fHI_V$, requires assumptions on the volume of LLSs and DLAs, which are model-dependent. An example of such model, and the inferred value of $\fHI_V$, is discussed in \S~\ref{sec:web}.}.

The left panel of Fig.~$\ref{fig:HIcotent}$ further shows the mass-weighted $\HI$ fraction embedded in each type of $\LyA$ absorber. The dominant reservoir of $\HI$ gas is the high-column density $\LyA$ absorbers, mainly DLAs. The $\fHI_M\sim1\%$ embedded in DLAs stays approximately constant over $2<z<5$, while the $\HI$ fraction embedded in super-LLS and LLS, which is the second dominant $\HI$ gas reservoir, increases with redshift. The diffuse IGM, represented by the $\LyA$ forest absorbers, is highly ionized and remains a minor reservoir of neutral gas.

\subsection{Observational Constraints on $\HI$ During Reionization}
In the right panel of Fig.~\ref{fig:HIcotent} we have compiled various inferred values of the volume-weighed $\HI$ fraction available in the literature from CMB (\citealt{2014A&A...571A..16P,2015arXiv150201589P}), Gunn-Peterson optical depth (\citealt{2006AJ....132..117F}), dark pixels (\citealt{2011MNRAS.415.3237M}), Gamma Ray Burst afterglow (\citealt{2008MNRAS.388.1101M}; \citealt{2006PASJ...58..485T}), quasars (QSOs) near zone (\citealt{2011MNRAS.416L..70B}; \citealt{2013MNRAS.428.3058S}), $\LyA$ luminosity function, equivalent width distribution, $\LyA$ fraction, and correlation function (\citealt{2010ApJ...723..869O}; \citealt{2011MNRAS.414.2139D}; \citealt{2013MNRAS.428.1366J}; \citealt{2015MNRAS.446..566M}; \citealt{2015MNRAS.452..261C}). We also show our suggested constraint using the $\LyA$ luminosity function alone (faint orange box, see \S~\ref{sec:Lya_LF}) and when combined with the equivalent width distribution (darker orange box, see \S~\ref{sec:constraining}).

All the open points are simulation(model)-calibrated measurements, which use the $\LyA$ radiative transfer modelling in the IGM around galaxies and QSOs. While  previous works make very sensible assumptions to interpret the observed data, astrophysical systematics in such simulation(model)-calibrated measurements may raise questions about the robustness of the inferred values.  While the present estimates at $5<z<7$ favour a volume-weighted $\HI$ fraction as high as $\sim50\%$ if taken at face value, it should be kept in mind that these estimates are implicitly assuming a bubble model.

Interestingly, recent constraints (\citealt{2015MNRAS.446..566M}; \citealt{2015MNRAS.452..261C}) including both large-scale patchy reionization and small-scale absorbers still favour values for the $\HI$ fraction $\gtrsim40\%$ at $z>7$. Our work also prefers numbers in this range. It is the aim of this paper to understand the reason for this.

\subsection{Theoretical expectations for $\HI$}\label{sec:HImodel}
The goal of this section is to highlight the need for a hybrid web-bubble model to interpret high-$z$ galaxy observations. We present theoretical estimates of $\fHI_V$ using analytic models for the three different classes of ionization structure in the IGM. These calculations illustrate the redshift evolution of $\fHI_V$ in the web (\S~\ref{sec:web}), bubble (\S~\ref{sec:bubble}), and hybrid web-bubble (\S~\ref{sec:hybrid}) model.

\begin{figure}
\centering
  \includegraphics[angle=-90,width=0.5\textwidth]{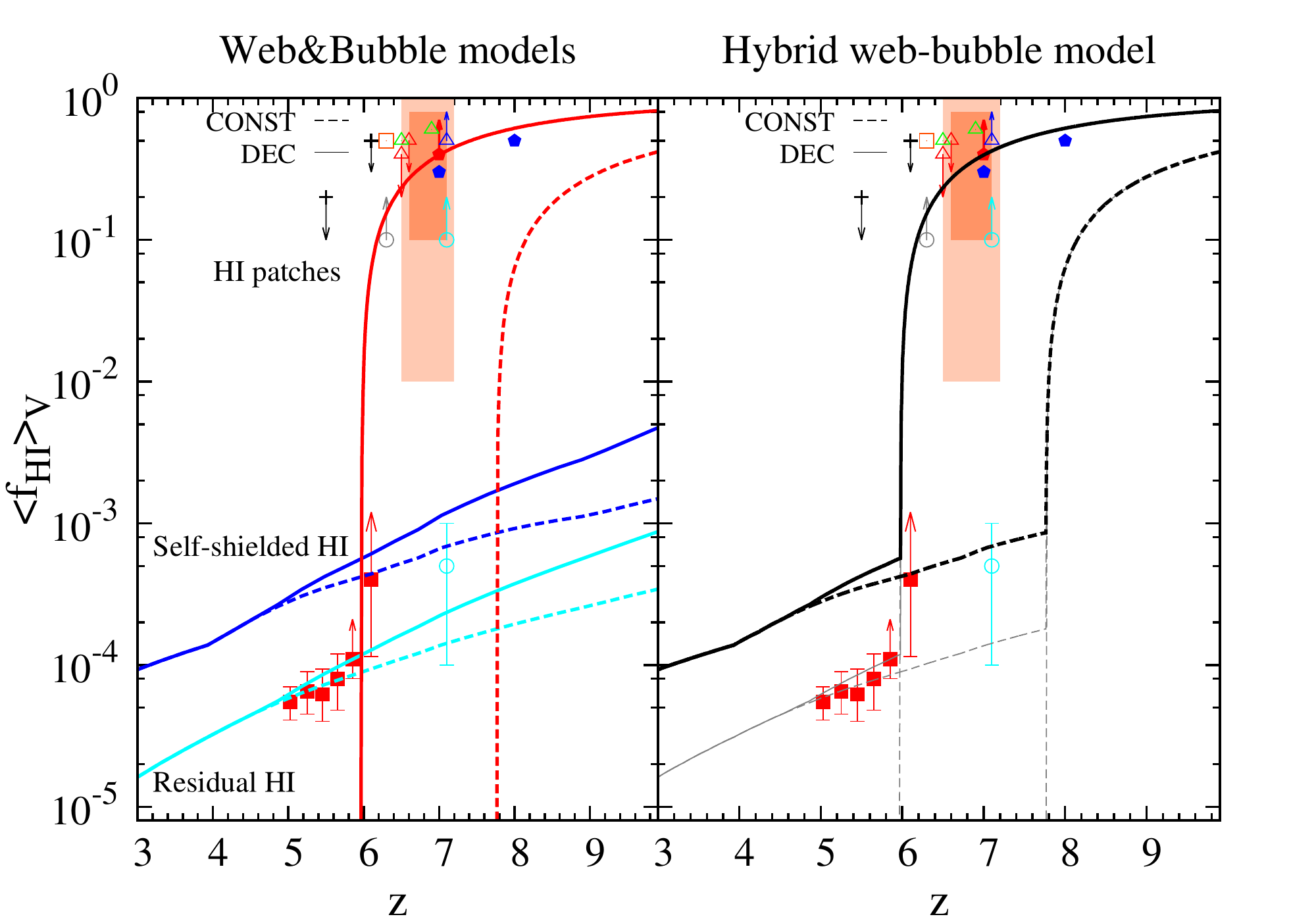}
  \caption{Redshift evolution of the volume-weighted $\HI$ fraction obtained with the analytic models. 
The blue and cyan lines refer to the residual neutral gas in the ionized regions (residual $\HI$, cyan) and self-shielded gas (self-shielded $\HI$, blue), i.e. the two different reservoirs in the web model. The red lines refer to the bubble model, where patchy neutral gas outside ionized bubbles ($\HI$ patches) is the main reservoir of $\HI$. Results for the web-bubble model are shown in the right panel for a case in which the residual $\HI$ inside ionized bubbles is calculated including only the diffuse (gray lines) or the self-shielded (black lines) gas. CONST and DEC models are indicated as dashed and solid lines. The points in Fig. $\ref{fig:HIcotent}$, inferred from observations, are overlaid as a guide, while the faint and dark orange boxes refer to our estimate based on the Ly$\alpha$ luminosity function alone and used in combination with the equivalent width distribution, respectively. This plot illustrates how hybrid models are required to explain the observational constraints at all redshifts.
}
   \label{fig:fHImodels}
\end{figure}

\subsubsection{$\HI$ Fraction in the Web Model}
\label{sec:web}
 In the web model, $\fHI_V$ is expected to increase with increasing redshift due to decreasing photoionization rate, and/or increasing mean gas density (by Hubble expansion).  $\fHI_V$ can be estimated as (e.g \citealt{2000ApJ...530....1M};  \citealt{2007MNRAS.382..325B})
\begin{equation}
\fHI_V=\int_0^{\Delta_{ss}}\xHI(\Delta_b)P(\Delta_b)d\Delta_b
      +\int_{\Delta_{ss}}^\infty P(\Delta_b)d\Delta_b,
\label{eq:fvweb}
\end{equation}
where $\Delta_b$ is the baryon overdensity, $P(\Delta_b)$ is the volume-weighted overdensity probability distribution function\footnote{$P(\Delta_b)$ is adopted from \cite*{2000ApJ...530....1M} with the redshift extrapolation of \cite{2004ApJ...601...64B}.}, and $\Delta_{ss}\propto\Gamma^{2/3}$ is the density threshold above which the gas self-shields against the UV background (\citealt{2001ApJ...559..507S}; \citealt{2005MNRAS.363.1031F}). $\xHI(\Delta_b)=\alpha_A(T)\bar{n}_H^{com}(1+z)^3f_e\Delta_b/\Gamma$ is the neutral fraction obtained assuming local photoionization equilibrium with a uniform photoionization rate $\Gamma$ (s$^{-1}$), $\bar{n}^{com}_H$ is the average comoving hydrogen number density, $\alpha_A$ is the case A recombination rate at temperature $T$, and $f_e$ is the electron fraction per hydrogen atom. The first and second term on the right hand side of equation (\ref{eq:fvweb}) are the volume-weighted $\HI$ fraction embedded in residual $\HI$ in the diffuse IGM and the self-shielded gas, respectively. 

We consider two models for the redshift evolution of the photoionization rate $\Gamma$: the CONST model assumes a constant  $\Gamma=\Gamma(z=4.75)$, while the DEC model assumes a photoionization rate decreasing with increasing redshift, i.e. $\Gamma(z)=\Gamma(z=4.75)[(1+z)/5.75]^{-1.5}$ (\citealt{2011MNRAS.412.2543C})\footnote{$\Gamma(z=4.75)$ is chosen to be consistent with the UV background measurement from the $\LyA$ forest, i.e. $\log_{10}\Gamma(z=4.75)/(10^{-12}{\rm s}^{-1})=-0.029^{+0.156}_{-0.147}$ (\citealt{2013MNRAS.436.1023B}).}.

The blue and cyan lines in Fig.~\ref{fig:fHImodels} show an example of the redshift evolution of $\fHI_V$, with the two different contributions from residual $\HI$ in the diffuse IGM (residual $\HI$; cyan) and neutral self-shielded gas (self-shielded $\HI$; blue). The global $\HI$ fraction is clearly dominated at all redshifts by the self-shielded gas.  While, as expected, the $\HI$ fraction increases with redshift due to the larger mean gas density, $\fHI_V$ increases more markedly in the DEC model due to the lower photoionization rate.

\subsubsection{$\HI$ Fraction in the Bubble Model}
\label{sec:bubble}
In the bubble model, the time evolution of the volume filling factor of ionized bubbles, the `porosity' factor $Q_i$, is given by (e.g. \citealt*{1999ApJ...514..648M})
\begin{equation}
\frac{dQ_i(t)}{dt}=\frac{\dot{n}_{ion}(t)}{\bar{n}_H^{com}}-\frac{Q_i(t)}{\bar{t}_{rec,B}(t)}, \label{porosity_eq}
\end{equation}
where $\dot{n}_{ion}$ is the ionizing photon emissivity (in units of $\rm ph~s^{-1}~cm^{-3}$) and $\bar{t}_{rec,B}=[\alpha_B\bar{n}^{com}_H(1+z)^3C]^{-1}$ is the case B recombination timescale with clumping factor $C$. If `empty' bubbles are assumed, we have
\begin{equation}
\fHI_V=1-Q_i.
\end{equation}

As with the web model, we consider two cases: the CONST model assumes a constant ionizing photon emissivity\footnote{$\dot{n}_{ion}(z)$ is anchored at $z=4.75$ based on $\LyA$ forest constraints, indicating $-0.46\lesssim\log_{10}\dot{n}_{ion}/10^{51}\rm{ph~s^{-1}cMpc^{-3}}\lesssim0.35$ over $2.40\lesssim z \lesssim 4.75$ (\citealt{2013MNRAS.436.1023B}). Here we assume $\dot{n}_{ion}(z=4.75)=10^{51}\rm{ph~s^{-1}cMpc^{-3}}$.} $\dot{n}_{ion}(z)=\dot{n}_{ion}(z=4.75)$, while the DEC model assumes that the ionizing photon emissivity decreases with increasing redshift, i.e. $\dot{n}_{ion}(z)=\dot{n}_{ion}(z=4.75)[(1+z)/5.75]^{-1.5}$. This choice of redshift evolution is made to bracket the possible range of parameters satisfying the \citet{2006AJ....132..117F} constraints.

In the left panel of Fig.~\ref{fig:fHImodels} the redshift evolution of $\fHI_V$ in the bubble model (red lines) shows a rapid change at $z \sim 6-8$, when $\fHI_V$ plummets to zero once reionization ends. A smooth transition to the post-reionized IGM, where small-scale absorbers must exist, is clearly  absent from these models as no $\HI$ gas is present inside ionized bubbles.
The behaviour in the CONST and DEC cases is very similar, with an earlier reionization in the former case, where a larger photoionization rate is present.

\subsubsection{$\HI$ Fraction in the Web-Bubble Model}\label{sec:hybrid}

In the web-bubble model we assume that ({\it i}) gas inside ionized bubbles behaves as in the web model, and ({\it ii}) gas outside ionized bubbles is fully neutral. These assumptions lead to 
\begin{equation}
\fHI_V=1-\left[1-\int_0^{\Delta_{ss}}\xHI(\Delta_b)P(\Delta_b)d\Delta_b
      -\int_{\Delta_{ss}}^\infty P(\Delta_b)d\Delta_b\right] Q_i, \label{hybrid}
\end{equation}
where the terms in square brackets are the $\HII$ fraction inside the ionized bubbles.

The redshift evolution of $\fHI_V$ in the web-bubble model is shown in the right panel of Fig. \ref{fig:fHImodels} for a case in which the residual $\HI$ inside ionized bubbles is calculated including only the diffuse (gray lines) or the self-shielded (black lines) gas. The assumed values of the photoionization rate and ionizing photon emissivity are the same as used in the previous sections. 
A comparison between the left and right panels of the figure shows that the web-bubble model produces a smooth transition from the bubble model (patchy reionization) during the EoR to the web model (dominated by small-scale absorbers) in post-reionization. 

Hence, to coherently explain and interpret present observations, a unified framework that includes both large-scale bubbles {\it and} small-scale absorbers is essential because ({\it i}) the presence of small-scale absorbers at lower-$z$ is evident from observations (Fig. \ref{fig:HIcotent}), and ({\it ii}) a smooth transition from a patchy reionization to a post-reionized IGM with small-scale absorbers is only possible within a hybrid web-bubble model (Fig. \ref{fig:fHImodels}).

\section{Cosmological $\LyA$ radiative transfer}\label{analytic}

In this section we present the formalism adopted to follow the cosmological $\LyA$ transfer through the reionization models discussed above. 

The general equation describing line transfer in the Lagrangian fluid frame is (\citealt{1984oup..book.....M}; \citealt{2004rahy.book.....C};
\citealt{2009RvMP...81.1405M}; \citealt{2014PASA...31...40D})
\begin{align}
&\frac{1}{c}\frac{\partial I_{\nu}}{\partial t}
+\boldsymbol{n}\cdot\boldsymbol\nabla I_{\nu}
-\frac{H+\boldsymbol{n\cdot\nabla v\cdot n}}{c}\nu\frac{\partial I_{\nu}}{\partial \nu}+
3\frac{H}{c}I_{\nu}\nonumber\\
&~~
=-\sigma_\alpha\nHI\varphi_\nu I_{\nu}+\sigma_\alpha \nHI\int\mathcal{R}(\nu,\nu')J_\nu(\nu')d\nu'+\varepsilon_\nu,
\label{EqLyART}
\end{align}
where $I_\nu$ is the specific intensity, $J_\nu$ is the angle-averaged intensity,
$\varepsilon_\nu$ is the $\LyA$ emissivity, $\boldsymbol{v}$ is the peculiar velocity, $\boldsymbol{n}$ 
is the unit direction vector of rays, $\sigma_\alpha=0.011$~cm$^2$Hz is the $\LyA$ cross section, and
$\varphi_{\nu}$ is the line profile of the $\LyA$ resonance line (units Hz$^{-1}$). 
The $\boldsymbol{n\cdot\nabla v\cdot n}$ term is the Doppler shift effect and 
$\mathcal{R}(\nu,\nu')$ is the redistribution function describing 
the resonant scattering of $\LyA$ photons.

There are generally no analytic solutions to equation~(\ref{EqLyART}). However, by performing a separation of scales, the problem can be simplified: multiple scattering effects are predominant on ISM scales because the surface brightness of $\LyA$ photons that are scattered back into the line-of-sight at IGM scales is typically negligibly small.  As scatterings on such small ISM scales can be effectively treated as a modification of the intrinsic line profile, and the scattering term can be overall neglected ({\citealt*{2011ApJ...728...52L}).

Equation (\ref{EqLyART}) can then be readily integrated along a line-of-sight to give the so-called `$e^{-\tau}$ approximation' (e.g. \citealt{1996ApJ...461...20H}; \citealt{2009RvMP...81.1405M}; \citealt{2007MNRAS.381...75M}). In this approximation, the $\LyA$ flux $F_\alpha$ observed from a $\LyA$-emitting galaxy at redshift $z_s$ is given by 
\begin{equation}
F_\alpha=\frac{L_\alpha}{4\pi D_L^2(z_s)}\int S_\nu(\nu_e)e^{-\tau_\alpha(\nu_e)} d\nu_e
=\frac{L_\alpha \mathcal{T}_{IGM}}{4\pi D_L^2(z_s)},
\end{equation}
where $\nu_e$ is the frequency of the Ly$\alpha$ photon when it is emitted, $D_L(z_s)$ is the luminosity distance, $L_\alpha$ is the intrinsic bolometric $\LyA$ luminosity (in units of erg~s$^{-1}$), $S_\nu(\nu_e)$ (in units of $\rm{Hz}^{-1}$) is the effective intrinsic line profile (including the effect of the ISM/CGM) normalized such that $\int S_\nu(\nu_e)d\nu_e=1$. $\mathcal{T}_{IGM}=\int S_\nu(\nu_e)e^{-\tau_\alpha(\nu_e)}d\nu_e$ denotes the IGM transmission factor (\citealt*{2011MNRAS.414.2139D}), and the $\LyA$ optical depth $\tau_\alpha(\nu_e)$ is
\begin{equation}
\tau_\alpha(\nu_e)\approx\sigma_\alpha\int_0^{l_p}dl'_p
\nHI(l'_p)\varphi_\nu\left[T,\nu_e\left(1-\frac{v_{tot}(l'_p)}{c}\right)\right],
\label{LyaOptDpt}
\end{equation}
where $T$ is the gas temperature and $v_{tot}=H(z_s)l_p+v(l_p)$ is the sum of the Hubble flow and the peculiar velocity. It is customary to express $\nu_e$ in terms of a velocity shift, i.e. $\Delta v/c=1-\nu_e/\nu_\alpha$. In the following we will use this convention. 

We would like to note here that by using the $e^{-\tau}$ approximation we ignore photons that scatter back into the line-of-sight, which would give rise to a low surface brightness `fuzz'. \cite{2011ApJ...728...52L} compared the $e^{-\tau}$ approximation to a full Monte-Carlo Ly$\alpha$ radiative transfer approach finding that the $e^{-\tau}$ approximation provides a good description of the transfer through the IGM as long as this is assumed to start at a distance larger than $1.5$ times the virial radius of the dark matter halo hosting a Ly$\alpha$ galaxy.
We have verified that this condition is met throughout our work.

We also introduce the mean IGM transmission factor and effective optical depth to characterize the typical impact of the intergalactic environment around $\LyA$-emitting galaxies. The mean $\LyA$ flux of many $\LyA$-emitting galaxies is $\langle F_\alpha\rangle\approx\langle L_\alpha\rangle \langle\mathcal{T}_{IGM}\rangle/(4\pi D_L^2)$,  where
\begin{equation}
\langle\mathcal{T}_{IGM}\rangle\approx
\int \langle S_\nu(\nu_e)\rangle e^{-\tau_\alpha^{\rm{eff}}(\nu_e)}d\nu_e
\end{equation}
is the mean IGM transmission factor and $\tau_\alpha^{\rm{eff}}=-\ln\langle e^{-\tau_\alpha}\rangle$ is the effective optical depth (e.g. \citealt{1996ApJ...461...20H}). Here we have assumed that the intrinsic line profiles of $\LyA$ galaxies and the optical depth of the IGM are uncorrelated, i.e. that $\langle S_\nu(\nu_e) e^{-\tau_\alpha(\nu_e)}\rangle\approx\langle S_\nu(\nu_e)\rangle e^{-\tau_\alpha^{\rm{eff}}(\nu_e)}$}.

The optical depth contribution from different intervening IGM absorbers (the diffuse neutral IGM outside ionized bubbles and the small-scale absorbers) is additive, i.e. $\tau_\alpha=\tau_{bub}+\tau_{web}$. The same applies to the effective optical depth, i.e. $\tau_\alpha^{\rm{eff}}=\tau_{bub}^{\rm{eff}}+\tau_{web}^{\rm{eff}}$. In the bubble [web] model of reionization we ignore $\tau_{web}$ [$\tau_{bub}$], while in the web-bubble model we include both. These two terms are discussed in more detail in the following sections.

\subsection{$\LyA$ Opacity from Large-Scale $\HI$ Patches}
\label{sec_bubble}

In the bubble model the $\LyA$ optical depth is due to diffuse expanding neutral IGM outside ionized bubbles ($\HI$ patches). The $\LyA$ optical depth of a homogeneous $\HI$ patch extending between comoving distance from a $\LyA$-emitting galaxy $R_1$ and $R_2$, can be written as (\citealt{2008MNRAS.385.1348M}; \citealt{2014PASA...31...40D})
\begin{equation}
\tau_{patch}(\nu_e,R_1,R_2)=\tau_{GP}\int_{x(\nu_e,R_2)}^{x(\nu_e,R_1)}\phi(x)dx,
\end{equation}
where $\phi=\Delta \nu_D\varphi_\nu$ is the dimensionless line profile, $\tau_{GP}=c\sigma_\alpha\bar{n}^{com}_{\rm{H}}(1+z_s)^3/(\nu_\alpha H)\approx4.44\times10^5[(1+z_s)/7.6]^{3/2}$ is the Gunn-Peterson optical depth, and $x(\nu_e,R)=\{\nu_e(1-H(z_s)R/[(1+z_s)c])-\nu_\alpha\}/\Delta \nu_D$. $\Delta \nu_D=\frac{\nu_\alpha}{c}\sqrt{\frac{2k_BT}{m_p}}$ is the Doppler width, with $k_B$ Boltzmann constant and $m_p$ proton mass.

In general, the $\LyA$ optical depth along a line-of-sight in the bubble model is given by:
\begin{equation}
\tau_{bub}(\nu_e)=\sum_{i=1}\tau_{patch}(\nu_e,R_{1,i},R_{2,i}),
\end{equation}
where $R_{1,i}$ [$R_{2,i}$] is the near [far] side of the edge of the $i$-th $\HI$ patch. The effective optical depth through an ensemble of $\HI$ patches is 
\begin{equation}
e^{-\tau_{bub}^{\rm{eff}}(\nu_e) }=\int e^{-\tau_{bub}(\nu_e)}P[\tau_{bub}(\nu_e)]d\tau_{bub}(\nu_e),
\end{equation}
where $P[\tau_{bub}(\nu_e)]$ denotes the probability distribution for $\tau_{bub}(\nu_e)$, which must be obtained from cosmological realizations of the bubble model.

There is a simpler limiting analytic case if we assume an ensemble of single large $\HI$ patches along all lines-of-sight. In the limit of a large $\HI$ patch ($R_2\rightarrow\infty$), the optical depth along a line-of-sight is
$\tau_{patch}(\nu_e,R_1)\approx\frac{\tau_{GP}\Lambda}{4\pi^2\nu_\alpha}
\left|\frac{\nu_e}{\nu_\alpha}\left[1-\frac{H(z_s)R_1}{(1+z_s)c}\right]-1\right|^{-1}$,  where $\Lambda=6.25\times10^8$~s$^{-1}$ is the damping coefficient (e.g. \citealt{1998ApJ...501...15M}; \citealt{2008MNRAS.388.1101M}). Then, for an ensemble of large $\HI$ patches we can evaluate the effective optical depth as 
\begin{equation}
e^{-\tau_{bub}^{\rm{eff}}(\nu_e) }\approx\int e^{-\tau_{patch}(\nu_e,R_1)}P(R_1)dR_1,
\label{eq:analytic_bub}
\end{equation}
where $P(R_1)dR_1$ is the probability to find the near side of a $\HI$ patch at a distance $R_1$ from a $\LyA$-emitting galaxy (for a related definition of bubble size distribution, see \citealt{2007ApJ...669..663M}). We model $P(R_1)$ as a Schechter function, $P(R_1)\propto R^{\alpha_1}\exp(-R_1/R_\ast)$, normalized as $\int P(R_1)dR_1=1$; $\alpha_1$ and $R_\ast$ are free parameters. We compare this analytic estimate of the effective optical depth to numerical calculations in \S~\ref{sec:LyA_RT_bub}.

\subsection{$\LyA$ Opacity from Small-Scale Absorbers}
\label{LyA_RT_webtopo}

In the web model  the $\HI$ gas is distributed in a collection of self-shielded absorbers. Each absorber is characterized by its $\HI$ column density, $\NHI$, and its proper velocity, $v_c$, relative to a given $\LyA$-emitting galaxy. The $\LyA$ optical depth through a single absorber is
\begin{equation}
\tau_{abs}(\nu_e)=
\sigma_\alpha\NHI\varphi_{\nu} \left[T_c,\nu_e\left(1-\frac{v_c}{c}\right)\right],
\label{cloud_optdpt}
\end{equation}
where $T_c$ denotes the gas temperature of an absorber. 

We introduce a novel analytic model of the $\LyA$ opacity from small-scale absorbers as follows. The effective optical depth of an ensemble of $\HI$ absorbers surrounding a $\LyA$-emitting galaxy is (see Appendix~\ref{app:tauweb} for a derivation)
\begin{align}
\label{eqtauweb}
\tau_{web}^{\rm{eff}}(\nu_e)&=\int d\NHI\frac{\partial^2\mathcal{N}}{\partial\NHI\partial z}\left|\frac{dz}{dl_p}\right|\times \\
&~~~~\int\frac{dv_c}{H(z_s)}\left[1+\xi_v(v_c,\NHI)\right]\left[1-e^{-\tau_{abs}(\nu_e,v_c,\NHI)}\right],
\nonumber
\end{align}
where $\xi_v(v_c,\NHI)$ is the galaxy-absorber correlation function in velocity space. We refer to a Gaussian streaming model (GSM) for $\xi_v(v_c,\NHI)$ when
\begin{align}
&1+\xi_v(v_c,\NHI)=\nonumber \\
&\int\frac{aHdr_{12}}{\sqrt{2\pi\sigma^2_{12}(r_{12})}}\left(1+\xi(r_{12})\right)\exp\left[-\frac{(v_c-aHr_{12}-\langle v_{12}(r_{12})\rangle)^2}{2\sigma_{12}^2(r_{12})}\right],
\label{xi_Gaussian}
\end{align}
where $r_{12}$ is the comoving separation between a galaxy and an absorber, $\xi(r_{12})$ is the real-space galaxy-absorber correlation function, $\langle v_{12}(r_{12})\rangle$ [$\sigma_{12}(r_{12})$] is the mean radial pairwise velocity [the pairwise velocity dispersion] between galaxy-absorber pairs, and $a=(1+z_s)^{-1}$ is the scale factor.

\subsubsection{The Region of Influence}
As the optical depth depends on $v_c$, it is useful to calculate the `critical' velocity, $v_{\rm crit}$, at which the optical depth of an absorber to a $\LyA$ photon emitted at frequency $\nu_e$ becomes unity for a given $\HI$ column density, i.e. $\tau_{abs}(v_c=v_{\rm crit})=1$. In fact, to first order, the $\LyA$ visibility is only affected by small-scale absorbers moving away from a central $\LyA$-emitting galaxy with $v_c<v_{\rm crit}$. We refer to the region that contains these absorbers as the `region of influence'. For high-column density absorbers such as LLS/DLA, the above condition is met in the wing of the absorption line profile $\varphi_\nu\approx \Lambda \left[4\pi^2(\nu_e(1-v_c/c)-\nu_\alpha)^2\right]^{-1}$. From  the Lorentz wing it follows that  for an absorber with $\HI$ column density $\NHI$,
\begin{equation}
\frac{v_{\rm crit}}{c}=
1-\frac{\nu_\alpha}{\nu_e}\left(1-\sqrt{\frac{\sigma_\alpha \NHI \Lambda}{4\pi^2\nu_\alpha^2}}\right).
\end{equation}  
If we set $\nu_e=\nu_\alpha$, then $v_{\rm crit}=c\sqrt{\frac{\sigma_\alpha \NHI \Lambda}{4\pi^2\nu_\alpha^2}}=507.3(\NHI/10^{20}{\rm cm}^{-2})^{1/2}$~km~s$^{-1}$. For a pure Hubble flow, the critical velocity corresponds to the comoving distance
\begin{equation}
D_{\rm{infl}}=\frac{v_{\rm crit}}{H_0}\frac{1+z}{[\Omega_m(1+z)^3+\Omega_\Lambda]^{1/2}}.
\end{equation}
As a reference, $D_{\rm{infl}}=3.5(\NHI/10^{20}{\rm cm}^{-2})^{1/2}h^{-1}$cMpc at $z=7$.

\vspace{1cm}
Armed with the analytic framework of $\LyA$ transfer described above to aid the understanding of our results, in the next section we perform cosmological hydrodynamical, radiative transfer simulations and derive a mock survey of $\LyA$-emitting galaxies with various reionization models.

\section{Simulations}\label{sec:sims}

In this section we describe the simulations used to model the observability of high redshift $\LyA$-emitting galaxies, and the mock galaxy catalogue obtained from them.

\subsection{Hydrodynamical Simulations of the IGM}

We employ the AMR cosmological N-body/hydrodynamical code RAMSES (\citealt{2002A&A...385..337T}) to simulate the IGM in a box of size  $25h^{-1}\rm cMpc$ on a side. The cosmological parameters are $(\Omega_m,\Omega_\Lambda,\Omega_b,\sigma_8,n_s,h)=(0.26,0.74,0.044,0.85,0.95,0.72)$. The total number of dark matter particles is $256^3$ with mass resolution of $m_{\scriptscriptstyle\rm{DM}}=5.5\times10^7h^{-1}\rm M_\odot$, and the gas is included on the $256^3$ base grid ($97.7h^{-1}$ckpc cell size) with two levels of refinement, reaching a $1024^3$ grid ($24.4h^{-1}$ckpc cell size) at the highest AMR refinement level. For our choice of the simulation parameters, the cosmological Jeans length is $\sim57({T/100{\rm K}})^{1/2}h^{-1}\rm{ckpc}$, which corresponds to a Jeans mass of $\sim 1.3\times10^7h^{-1}$M$_\odot$ at $T=100$~K. The finest spatial resolution therefore fulfills the minimum requirement to resolve self-shielded absorbers of order of the Jeans length (\citealt{2001ApJ...559..507S}).

The initial conditions are generated with the COSMICS package (\citealt{1995astro.ph..6070B}). The initial temperature is set to 650~K. This is higher than the value expected from the cooling and heating of the IGM after recombination as computed from RECFAST (\citealt*{1999ApJ...523L...1S}) to take into account the fact that our calculation did not include primordial heating mechanisms such as Compton heating by the CMB. The temperature is then calculated according to an adiabatic evolution. The initial redshift of the simulation is chosen as $z_{ini}=225$, to allow sufficient nonlinearity to develop at the reionization epoch $z\sim20-5$. 

The dark matter haloes are identified using the HOP algorithm (\citealt{1998ApJ...498..137E}) as implemented in the RAMSES package.

\subsection{Radiative Transfer Simulations}

\begin{table}
\centering
\caption{List of reionization models. The columns indicate, from left to right, the model and its name, 
the photoionization rate in terms of $\Gamma_{-12}=\Gamma/10^{-12}$~s$^{-1}$ as assumed at $z=6$ [7] in the bubble [web] model, and the resulting volume-weighted $\HI$ fraction, $\fHI_V$, at $z=7$.}
\label{table:rein}
    \begin{tabular}{lll}
    \hline\hline
    bubble model & $\Gamma_{-12}$($z$=6)	& $\fHI_V$($z$=7) \\
    \hline
    B1   			& 0.380	        & 0.365	    \\  
    B2 		        & 0.190	        & 0.676	    \\  
    B3		        & 0.019	        & 0.990	    \\
    \hline
    web model & $\Gamma_{-12}$($z$=7) & $\fHI_V$($z$=7)  \\
    \hline
    W1 	            & 0.1				& 0.00073 \\
    W2  	        & 0.01	        	& 0.012 \\  
    W3              & 0.005         	& 0.032 \\
    \hline
    web-bubble model & 				 	&  $\fHI_V$($z$=7)  \\
    \hline
    B1+W2           &          			& 0.373 \\    
    B1+W3           & 	       			& 0.387 \\    
    B2+W2           & 					& 0.681 \\
    B2+W3  	        & 	        		& 0.688 \\  
	\hline\hline   
    \end{tabular}
\end{table}

We use a two-way approach to follow the radiative transfer. In the first, we generate bubble models by performing full radiative transfer simulations of ionizing UV photons by post-processing the base AMR grid of the cosmological hydrodynamical simulation (\S~\ref{sim:bubble}). In the second, we generate web models by post-processing the finest AMR grid without full RT, but assuming photoionization equilibrium with a uniform UV background and a self-shielding criterion (\S~\ref{sim:web}). To generate the web-bubble models we modify the bubble models by treating  the regions inside the ionized bubbles as web models (\S~\ref{sim:web-bubble}). This approach enables us to access spatial scales for the self-shielding gas which are presently beyond the computational capability of the full radiative transfer simulations.

	We emphasize that the purpose of these simulations is not to produce the best possible reionization model, but to explore the impact of large-scale patchy reionization features (i.e. ionized bubbles) and small-scale absorbers on the observability of $\LyA$-emitting galaxies and on the inference of $\fHI_V$ using $\LyA$ surveys.

\begin{figure*}
  \centering
  \includegraphics[angle=0,width=\textwidth]{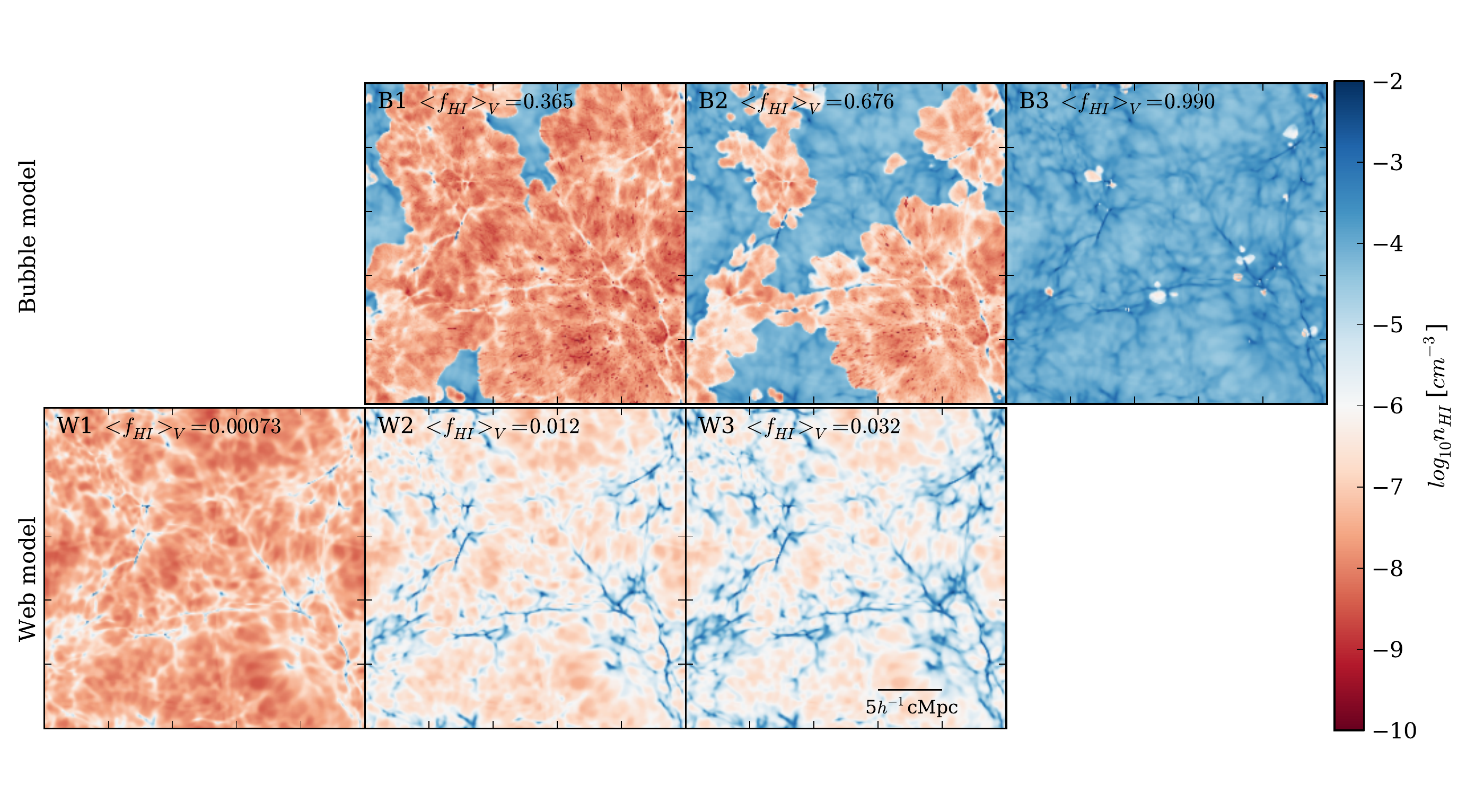}
  \caption{Neutral hydrogen number density, $\nHI$, at $z=7$ in slices of our simulations for the bubble (B1, B2 and B3; {\it top panels}) and web (W1, W2 and W3; {\it bottom panels}) models detailed in Table~\ref{table:rein}. Each snapshot is a x-y slice at $12.5h^{-1}$cMpc with $97.7h^{-1}$ckpc thickness. Panels in the same column give a similar suppression of the $\LyA$ visibility in the observed $\LyA$ luminosity functions  shown in Fig.~$\ref{fig:LyAlumfunc}$.}	
  \label{fig:reiontopology1}
  \includegraphics[angle=0,width=\textwidth]{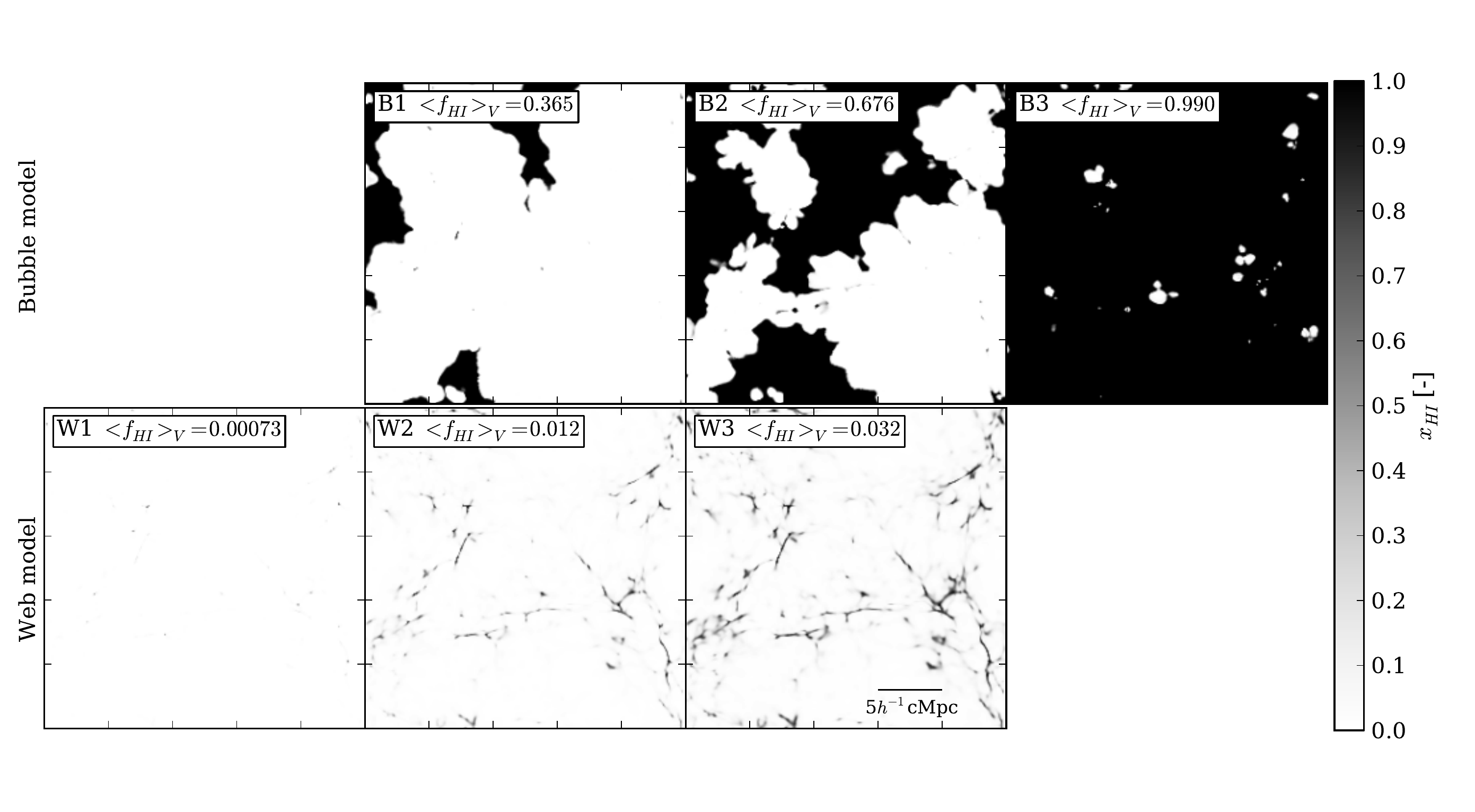}
  \caption{Same as in Fig. \ref{fig:reiontopology1} but for the local $\HI$ fraction $\xHI$.}
   \label{fig:reiontopology2}
\end{figure*}

\begin{figure*}
\centering
  \includegraphics[angle=0,width=\textwidth]{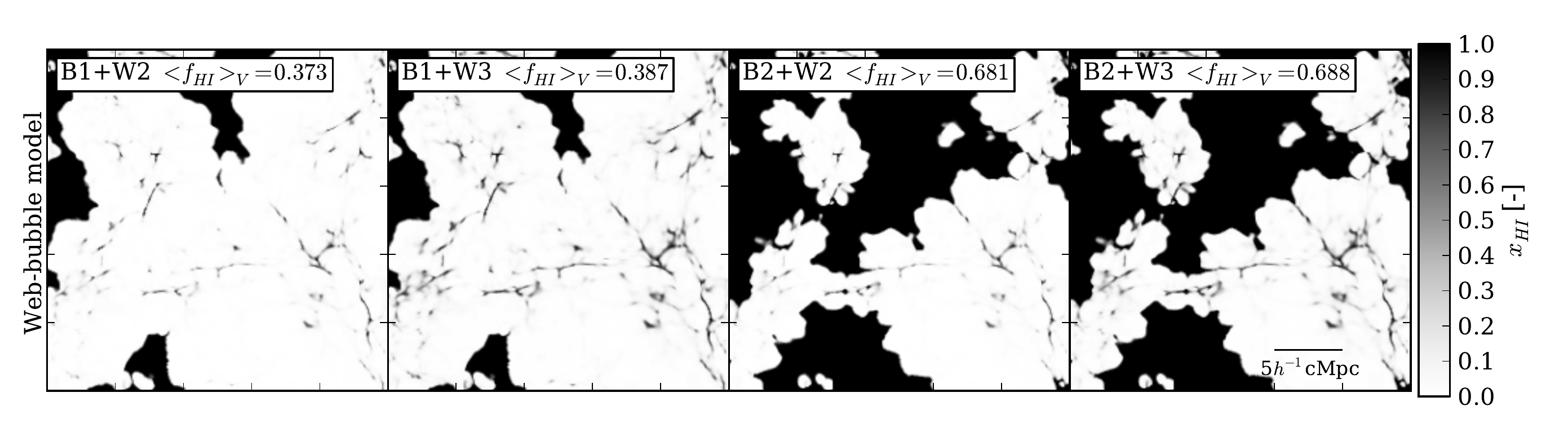}
  \caption{Slice map of the local $\HI$ fraction for hybrid web-bubble models.}
\label{fig:adhoc_hybrid_topology}
\end{figure*}

\subsubsection{Bubble Models}\label{sim:bubble}

We use the cosmological radiative transfer code CRASH (\citealt{2001MNRAS.324..381C}; \citealt{2003MNRAS.345..379M}; \citealt{2009MNRAS.393..171M}; \citealt{2013MNRAS.431..722G}) to generate our bubble models. For the full RT computation, we post-process the density and temperature fields on the $256^3$ cells of the hydrodynamical simulation. While our box size is not sufficient to include the largest ionized bubbles present during the later stages of reionization, this does not affect the goal of the paper.

The model for the ionizing sources is based on the one described in \cite{2012MNRAS.423..558C}:  the volume averaged ionizing emissivity, $\dot{n}_{ion}$ (photons s$^{-1}$ cMpc$^{-3}$), at $z>6$ is parameterized as $\dot{n}_{ion}(z)=10^{50.89}\chi(z)\frac{\alpha_b+3}{2\alpha}\left(\frac{\Gamma_{-12}(z=6)}{0.19}\right)$, where $\chi(z)=ae^{b(z-9)}\left[a-b+be^{a(z-9)}\right]^{-1}$, with $a=14/15$ and $b=2/3$ (see \citealt{2007MNRAS.382..325B}). The values of $\Gamma_{-12}(z=6)$ are shown in Table $\ref{table:rein}$. We assume that the ionizing emissivity is produced by galaxies with a power-law spectrum of slope $\alpha_b=\alpha=3$, and we distribute it among all haloes proportionally to their mass. 
 
We ran the radiative transfer simulation using 10 gas density and temperature snapshots from $z=15$ to $z=5$ equally spaced in redshift, including both hydrogen and helium with a number fraction 0.92 and 0.08, respectively. For each source, we emit $10^6$ photon packets distributed according to the power-law spectrum with 29 frequency bins sampled from 13.6eV to 200eV.

Finally, we produce a catalogue of bubble models for different values of $\Gamma_{-12}(z=6)$. Slices through these models are shown in the top panels of Figs.~\ref{fig:reiontopology1} and \ref{fig:reiontopology2}. 
In Fig.~\ref{fig:reiontopology1} the maps of $\HI$ number density clearly show that the global $\HI$ fraction increases as $\Gamma_{-12}(z=6)$ decreases (from left to right), as expected. More specifically, a volume-weighted $\HI$ fraction of $\fHI_V=0.365$, 0.676 and 0.990 is obtained at $z=7$ for $\Gamma_{-12}(z=6)=0.380$, 0.190 and 0.019, respectively. Furthermore, the figures show the characteristic feature of patchy reionization, i.e. large-scale bubbles.

Since there is a one-to-one correspondence between $\Gamma_{-12}(z=6)$ and $\fHI_V(z=7)$, we will use them interchangeably to specify the model. 

\subsubsection{Web Models}\label{sim:web}

We use the prescription of \cite{2013MNRAS.430.2427R} to account for self-shielding gas in the web models, which consists of a fitting function matched to their full RT transfer simulation. This prescription assumes photoionization equilibrium in each cell of the simulation with a modified background (see below), i.e.
\begin{equation}
\alpha_A(T)\nHII n_e=\Gamma_{Rah}\nHI,
\end{equation} 
where $n_e$ is the electron number density, and $\Gamma_{Rah}$ is the modified local photoionization rate. The neutral fraction in each cell is then given by $\xHI=\left[(\gamma+2)-\sqrt{(\gamma+2)^2-4}\right]/2$, where $\gamma=\Gamma_{Rah}/(\alpha_A \nH f_e)$. The factor $f_e=n_e/\nHII$ is 1 for a pure hydrogen medium, while $f_e>1$ if helium is included. We assume that the IGM temperature  is $T \sim10^4$K due to photoionization heating. The modified local photoionization rate is given by
\begin{equation}
\frac{\Gamma_{Rah}}{\Gamma}=
0.98\left[1+\left(\frac{\nH}{n_{SS}}\right)^{1.64}\right]^{-2.28}
+0.02\left[1+\left(\frac{\nH}{n_{SS}}\right)\right]^{-0.84},
\end{equation}
where $n_{SS}$ is the density at which the gas starts to be self-shielded 
\begin{equation}
n_{SS}=6.73\times10^{-3}\Gamma_{-12}^{2/3}\left(\frac{T}{10^4{\rm K}}\right)^{0.17} {\rm cm}^{-3}.\label{sst}
\end{equation}

To compute $\nHI$ using the above prescription we use the density field of the finest AMR level $1024^3$ from the hydrodynamical simulation\footnote{As a comparison, we have also calculated $\nHI$ using a threshold method, in which all the cells with gas density above $n_{SS}$ are assumed to be fully neutral, otherwise the neutral fraction is computed assuming photoionization equilibrium with $\Gamma$ rather than $\Gamma_{Rah}$, i.e. $\alpha_A(T)\nHII n_e=\Gamma\nHI$. We note that while mapping between assumed photoionization rate and the abundance of small-scale absorbers changes depending on the prescription, as long as $\fHI_V$ embedded inside small-scale absorbers is similar, the result is insensitive to the self-shielding prescription. Hence, the quantity that more directly impact the observation of $\LyA$-emitting galaxies is the number density of small-scale absorbers rather than the photoionization rate.}. 

The values adopted for the uniform photoionization rate $\Gamma_{-12}$ are found in Table $\ref{table:rein}$. Slices through our web models are reported in the lower panels of Figs.~\ref{fig:reiontopology1} and \ref{fig:reiontopology2}. Similarly to the bubble models, the maps show a higher neutral fraction for decreasing photoionization rate. However, the ionization structure of the IGM is significantly different, as the neutral gas is concentrated in high density peaks where small-scale absorbers, whose distribution follows the structure of the cosmic web, reside\footnote{Note that web models are {\it not} equivalent to outside-in reionization scenarios. They simply show the region of the universe that is reionized early in an inside-out scenario, with residual self-shielded $\HI$.}.

\subsubsection{Web-Bubble Models}\label{sim:web-bubble}

We generate the web-bubble models at $z=7$ as follows. First, we take a full RT simulation used to generate the bubble models (B1 and B2). Then, we recalculate the local $\HI$ fraction inside the ionized bubbles according to the web model with a photoionization rate $\Gamma_{-12}(z=7)=0.01$ and $0.005$ (W2 and W3) on the finest AMR grid\footnote{While in principle the photoionization rate inside bubbles is not independent of bubble size, we take this as a convenient free  \textit{ad hoc} parameter to adjust the abundance of self-shielded absorbers inside bubbles.}. In practice, $\xHI$ is calculated locally as the maximum between the values obtained from the bubble and the web model.
Our web-bubble models are catalogued in Table \ref{table:rein}. 

Slices through the web-bubble models are shown in Fig.~\ref{fig:adhoc_hybrid_topology} in terms of $\xHI$ map. As expected, the evolution of $\xHI$ with photoionization rate is the same as the one in the web and bubble models. Quantitatively, though, the neutral fraction here is slightly higher than the one in the corresponding  bubble models due to the contribution of small-scale absorbers (see Table \ref{table:rein}). In addition, the ionization structure of the IGM looks like a combination of the one from the  bubble and web models, as the small-scale absorbers appear as impurities inside large-scale ionized bubbles. 

\begin{figure*}
\centering
  \includegraphics[angle=0,width=0.9\textwidth]{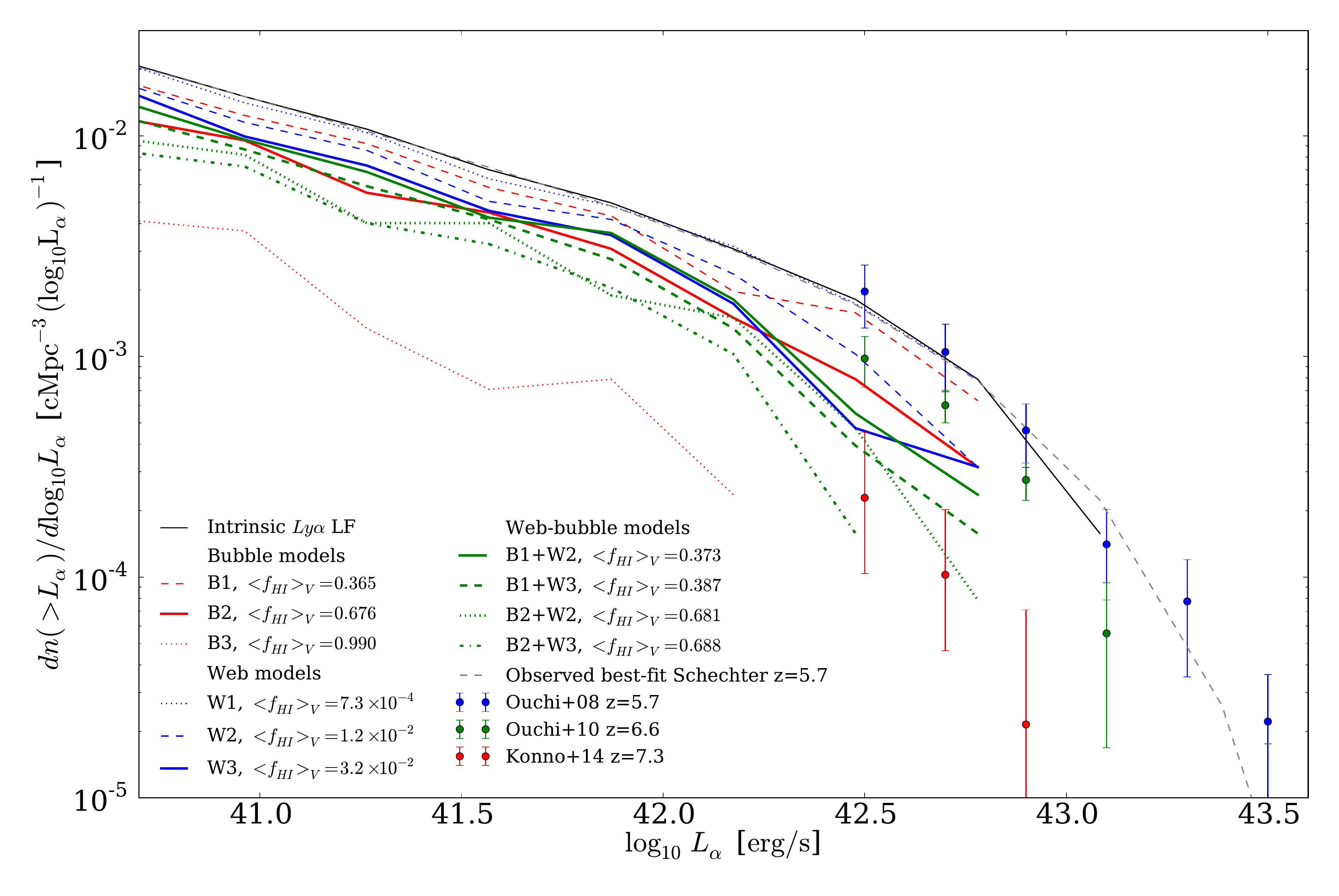}
  \caption{Intrinsic (black line) and observed differential $\LyA$ luminosity functions at $z=7$ as expected for bubble (red lines), web (blue lines), and web-bubble (green lines) reionization models. The observed data points of Ouchi et al. (2008, 2010) and Konno et al. (2014) are shown as blue, green, and red points. The best-fit Schechter function to the observations at $z=5.7$ is shown as the gray dashed line. For each model, the line style refers to a different value of the parameters, as indicated in Table \ref{table:rein}. The figure shows the degenerate impact of large-scale $\HI$ patches and small-scale absorbers on the $\LyA$ LF.
  }\label{fig:LyAlumfunc}
\end{figure*}

\subsection{Mock Galaxy Catalogue}\label{sec:mock}

The observed $\LyA$ luminosity of a galaxy is related to its intrinsic $\LyA$ luminosity via the IGM transmission factor $\mathcal{T}_{IGM}$ as $L_\alpha^{obs}=\mathcal{T}_{IGM}L_\alpha$ (this is discussed more in detail in \S~\ref{analytic}; see also Appendix \ref{appendix} for more technical aspects). We stress that  `intrinsic' here refers to the Ly$\alpha$ luminosity that a galaxy would have if the IGM were transparent. As our main results are insensitive to the precise model for the intrinsic luminosity function (LF), we only briefly describe the methodology applied to generate the intrinsic mock galaxy catalogue. 

We use the abundance matching technique (e.g. \citealt{2000MNRAS.318.1144P}; see also Appendix \ref{app:abundance_matching}) to populate dark matter haloes with Ly$\alpha$ emitting galaxies. We find the relation between halo mass and intrinsic Ly$\alpha$ luminosity by equating the {\it observed} cumulative $\LyA$ luminosity function $n(>L_\alpha)$ (in units of cMpc$^{-3}$) at $z=5.7$ (\citealt{2008ApJS..176..301O}) to the simulated halo mass function $dn(>M'_h)/dM'_h$ at $z=7$, 
\begin{equation}
n(>L_\alpha)=f_{duty}\int^{\infty}_{M_h(L_\alpha)}\frac{dn(>M'_h)}{dM'_h}dM'_h,
\end{equation}
where $f_{duty}$ is the duty cycle and $M_h(L_\alpha)$ is the halo mass corresponding to a $\LyA$ luminosity $L_{\alpha}$. We thus assume that the intrinsic Ly$\alpha$ luminosity function at $z=7$ is equal to the observed one at $z=5.7$, and that the difference between $z=5.7$ and $z=7$ is entirely due to the IGM. We therefore constrain the IGM opacity using the {\it variation} of the $\LyA$ LF relative to that in the post-reionization Universe
\footnote{The small-scale absorbers in the post-reionized universe may affect the visibility of $\LyA$-emitting galaxies at $z<6$. Hence, calibrating with $z<6$ $\LyA$ LF may not give a truly `intrinsic' $\LyA$ luminosity as defined above. This contribution should in principle be subtracted. However, as we will confirm in \S~\ref{sec:LyA_RT_web}, the impact of small-scale absorbers at $z\lesssim6$ is small.  Note, though, that ignoring the post-reionization optical depth of the IGM causes us to {\it underestimate} the intrinsic Ly$\alpha$ luminosity of galaxies, which in turn leads us to underestimate the IGM opacity and hence the neutral fraction in the IGM (see Dijkstra et al. 2011).}.
The abundance matching technique gives a semi-empirical relation between the halo mass and the intrinsic $\LyA$ luminosity for each $f_{duty}$ (examples are shown in Fig. \ref{fig:abundance_matching}). In our fiducial case we use $f_{duty}=1$. 
We then populate each halo with a single $\LyA$-emitting galaxy of intrinsic $\LyA$ luminosity given by the $M_h-L_{\alpha}$ relation. 

Because observations are available only down to $\log_{10} [L_{\alpha}/({\rm erg \, s}^{-1})]\approx 42.5$, to extend the calculations to lower luminosities we extrapolate assuming a faint-end slope of 1.5 (Ouchi et al. 2008, but see \citealt{Gronke15,Dressler15} for both theoretical and observational support for significantly steeper slopes of $\approx 2.2$). We note that, because of the small box size (which is needed to include small-scale absorbers and the faint galaxies responsible for reionization), the simulated LFs only extend to $\log_{10}[L_\alpha/({\rm erg \, s}^{-1})]\approx42.8$.

We model the $\LyA$ transfer in the ISM/CGM through the $\LyA$ spectral line profile (e.g. \citealt{2011MNRAS.414.2139D}; \citealt{2014MNRAS.444.2114J}; \citealt{2015MNRAS.452..261C}), by assuming a Gaussian profile with circular velocity $v_{circ}=20.4h^{1/3}(M_h/10^8{\rm M}_\odot)^{1/3}[(1+z)/7.6]^{1/2}$~km~s$^{-1}$ (\citealt{2004MNRAS.349.1137S}; \citealt{2010ApJ...716..574Z}), shifted redward by $\Delta v=600$~km~s$^{-1}$ to mimic the effect of scattering through a galactic wind (\citealt{2010MNRAS.408..352D}; \citealt{2011MNRAS.414.2139D}). This is rather arbitrary, but \cite{2010ApJ...717..289S} and \cite{2015ApJ...807..180W} justify a number between 200-800~km~s$^{-1}$. While the quantitative results are affected by this choice, the qualitative conclusions in this paper remain valid. We point out that our model assumes a universal line profile and shift, while a distribution is more likely. Since $\mathcal{T}_{IGM}$ is highest for redshifted Ly$\alpha$ lines, this can bias samples of Ly$\alpha$-selected galaxies to larger $\Delta v$.

\begin{figure*}
\centering
\advance\leftskip-1cm
  \includegraphics[angle=0,width=1.1\textwidth]{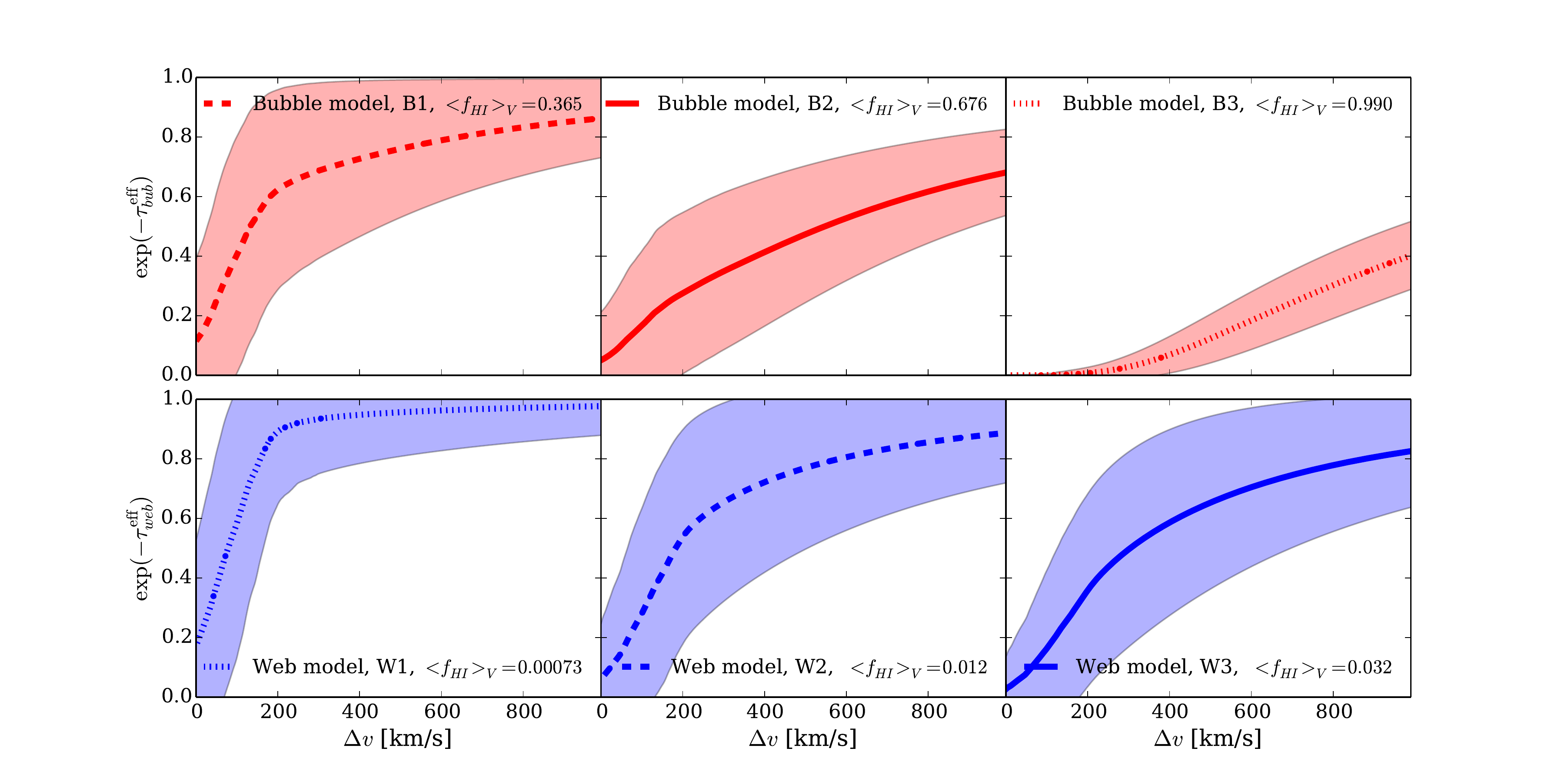}
  \caption{Effective optical depth (lines) extracted at $z=7$ from simulations of the bubble (B1, B2 and B3; {\it top panels}) and web (W1, W2 and W3; {\it bottom panels}) models. The shaded regions are the corresponding $1\sigma$ dispersion.
  }
   \label{fig:dampingwing_eff_HIpatch}
\end{figure*}

\section{Results}\label{sec:result}

\subsection{$\LyA$ Luminosity Function}\label{sec:Lya_LF}

We first show the impact of large-scale $\HI$ patches and small-scale absorbers on the $\LyA$ luminosity function in Fig. $\ref{fig:LyAlumfunc}$, which contains the  differential intrinsic $\LyA$ luminosity function of galaxies (black solid line) together with the predicted apparent luminosity function for our bubble (red lines), web (blue lines), and web-bubble (green lines) models with different values of $\fHI_V$. Fig.~\ref{fig:LyAlumfunc} shows that

\begin{itemize}
\item 
The predicted luminosity function decreases with $\fHI_V$ as naturally expected, because more neutral hydrogen in the universe increases the overall opacity to $\LyA$ photons. 

\item 
The relative abundance of large-scale bubbles and small-scale absorbers is a key factor to estimate the observed $\LyA$ luminosity function. The bubble, web, and web-bubble models predict almost identical luminosity functions for vastly different $\fHI_V$. For example, a bubble model with $\fHI_V=0.676$ (B2) gives rise to a luminosity function that is practically indistinguishable from that of a web model with $\fHI_V=0.032$ (W3) or of a web-bubble model with $\fHI_V=0.373$ (B1+W2). This was first pointed out by \cite{2013MNRAS.429.1695B}. 

\item
The presence of small-scale absorbers inside ionized bubbles provides an opacity additional to that from the neutral patches between large-scale bubbles. This is clear comparing e.g. the LFs from B1 (dashed red line) to those from B1+W2 (dotted green) or B1+W3 (dashed green).

\item Web models with $\fHI_V\sim 10^{-2}$ correspond to bubble models with $\fHI_V\sim 10^{-1}$. Table~\ref{table:rein} indicates that this requires $\Gamma \leq 10^{-14}$ s$^{-1}$. For example, the red dashed (B1) and blue dashed (W2) lines in Fig.~\ref{fig:LyAlumfunc} show that $\Gamma_{-12}(z=7)=0.01$ is needed for a web model to produce a LF similar to that of a bubble model with $\fHI_V\sim0.4$. This is in agreement with \cite{2015MNRAS.446..566M}.
\item Comparing the simulations to the observations of \cite{2008ApJS..176..301O,2010ApJ...723..869O} and \cite{2014ApJ...797...16K}, we conclude that at $z=6.6$, $40\% \lesssim \fHI_V \lesssim 70\%$ for the bubble model, $\fHI_V\sim1\%$ for the web model, and $\fHI_V\lesssim40\%$ for a web-bubble model. At $z=7$ we have instead $70\% \lesssim \fHI_V \lesssim 99\%$, $\fHI_V\gtrsim 3\%$, and $40\% \lesssim \fHI_V\lesssim70\%$, respectively.  The inferred $\HI$ fraction thus highly depends on the reionization model adopted.
\end{itemize}

While the aim of this paper is to present a proof of concept and we defer to future work more rigorous and precise constraints on the $\HI$ fraction, these results are in excellent agreement with existing work (\citealt{2013MNRAS.429.1695B}; \citealt{2015MNRAS.446..566M}; \citealt{2015MNRAS.452..261C}) and underline the importance of understanding the precise ionization structure of the IGM during the EoR in terms of both large-scale bubble features and small-scale absorbers. 
In the following, we use the simulations described in \S~\ref{sec:sims} and the analytic formalism outlined in \S~\ref{analytic} to gain more insight into the $\LyA$ RT and the inference of $\fHI_V$ from observed $\LyA$-emitting galaxies.

\subsection{The Red Damping Wing in Bubble Models}\label{sec:LyA_RT_bub}

We now analyse the $\LyA$ red damping wing opacity to quantify the impact of large-scale $\HI$ patches on the visibility of $\LyA$-emitting galaxies. The red lines in the top panel of Fig.~\ref{fig:dampingwing_eff_HIpatch} show the mean transmission $\exp(-\tau_{bub}^{\rm{eff}})$ as a function of $\Delta v$ for three different values of $\fHI_V$ (B1, B2, and B3 from left to right). We evaluate the effective optical depth directly as an average of $e^{-\tau_{bub}}$ using line-of-sight skewers from galaxies extracted from the simulations. The shaded region indicates the $1\sigma$ dispersion $\sigma_{\tau_{bub}}^2=\langle (e^{-\tau_{bub}}-e^{-\tau_{bub}^{\rm{eff}}})^2\rangle$. 
We have used 1185 lines-of-sight, i.e. equivalent to the number of galaxies in the simulation box.

The damping wing becomes more opaque with increasing neutral fraction and decreasing $\Delta v$. The opacity varies significantly between different lines-of-sight as indicated by the large dispersion of $\sigma_{\tau_{bub}}\sim 0.2$.

\subsubsection{Comparison to the Analytic Model}
\begin{figure}
\centering
  \includegraphics[angle=0,width=\columnwidth]{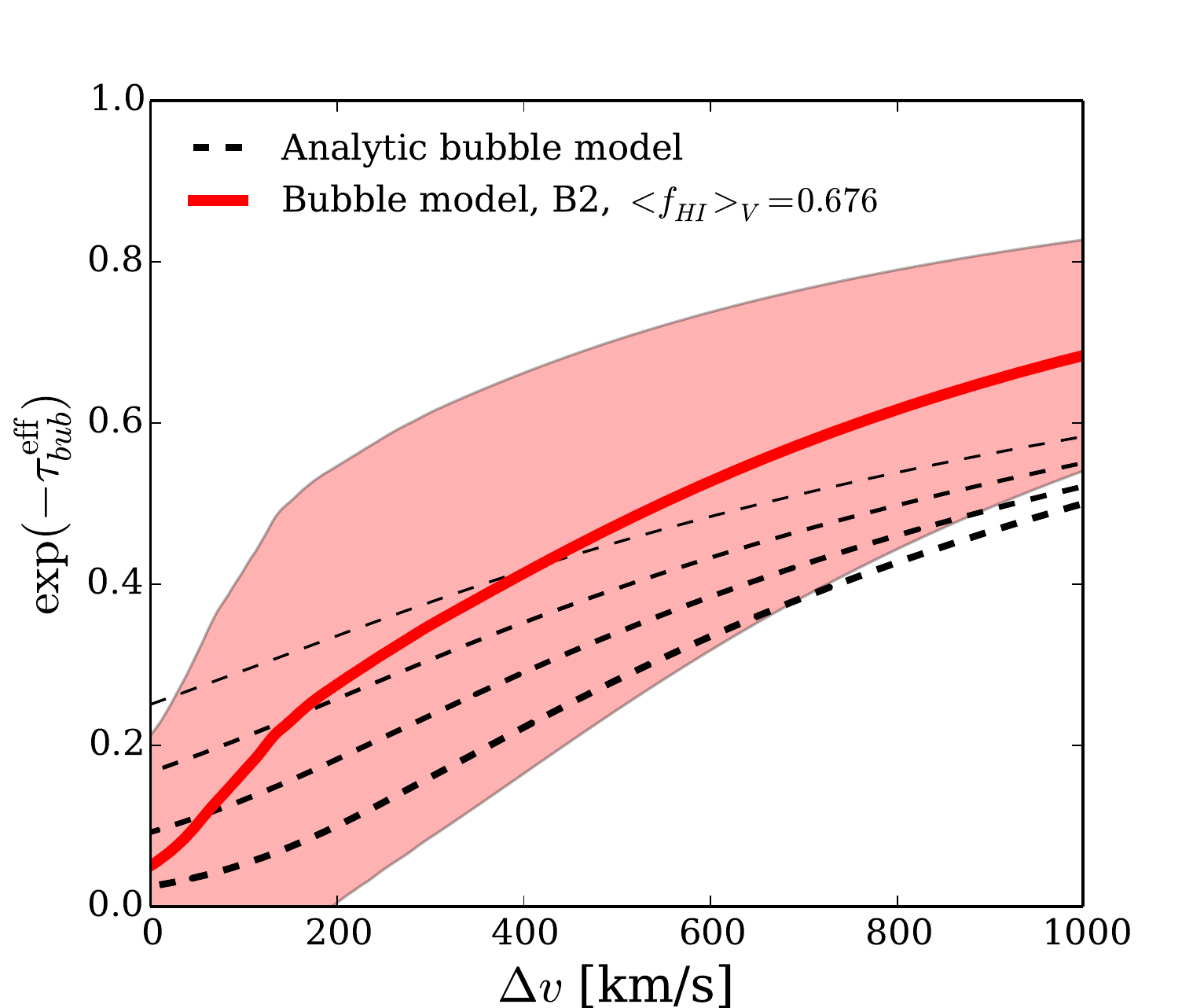}
  \caption{Comparison between the analytic and simulated bubble model. The black dashed lines are the effective optical depth from the analytic approximation (equation (\ref{eq:analytic_bub})) with a fixed slope $\alpha_1=0.5$ and $R_\ast=1.7,~3.0,~5.0,~10.0h^{-1}\rm cMpc$ (lines from bottom to top). The discrepancy between the simulation and the analytic model is due to the large single $\HI$ patch approximation used in the latter. See text for detail.}
  \label{fig:tau_eff_bubble_compare}
\end{figure}

\begin{figure}
\centering
  \includegraphics[angle=0,width=\columnwidth]{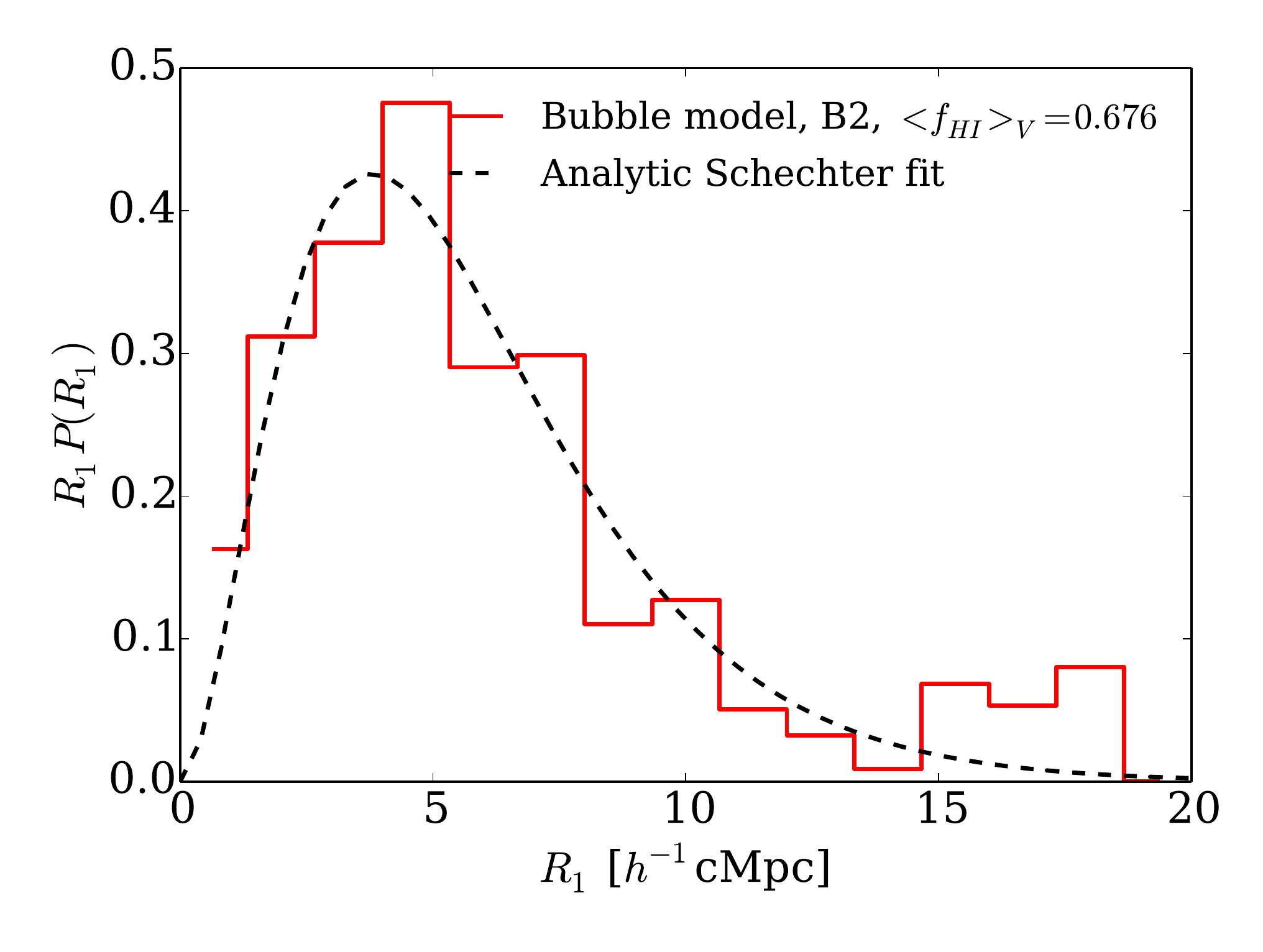}
  \caption{Distribution of the distance, $R_1$, to the near-side of $\HI$ patches from $\LyA$-emitting galaxies. The red line shows the result for the bubble model simulation B2. The black dashed line is the Schechter function fit to the simulated $P(R_1)$, where the best-fit parameters are $\alpha_1=0.49$ and $R_\ast=1.7h^{-1}\rm cMpc$.}
   \label{fig:bubble_size_PDF}
\end{figure}

To see how well the red damping wing can be captured by the analytic approximation, in Fig.~\ref{fig:tau_eff_bubble_compare} we compare the results from our B2 model to those obtained using equation~(\ref{eq:analytic_bub}) with $\alpha_1=0.5$ and $R_\ast=1.7,~3.0,~5.0,~10.0h^{-1}\rm cMpc$. Note that the case with $R_\ast=1.7$ (thickest black dashed line) represents the Schechter function fit to the simulated $P(R_1)$ distribution. Fig.~\ref{fig:bubble_size_PDF} shows that the Schechter function fit is indeed a good approximation to the simulation, in which the distance to the near-side of the closest $\HI$ patch peaks at $\sim 5h^{-1}$cMpc from a galaxy.

The comparison in Fig.~\ref{fig:tau_eff_bubble_compare} clearly indicates that the analytic model is too crude to capture the red damping wing behaviour found in the simulations, and systematically overestimates the optical depth, although the bubble size distribution is modelled reasonably well. The discrepancy highlights that the opacity is coming indeed from the neutral gas distributed among multiple $\HI$ patches, rather than in a single large $\HI$ patch, as assumed in the analytic model of equation~(\ref{eq:analytic_bub}). This in fact leads to an overestimate of the neutral gas and thus of the opacity.

In addition, the single large $\HI$ patch approximation is also responsible for a different shape of the damping wing, because the optical depth scales as $\Delta v^{-1}$ (e.g. \citealt{1998ApJ...501...15M}). On the other hand, the presence of multiple ionized bubbles in the simulations makes the medium more transparent, and hence the damping wing profile steeper. This implies that, unless the analytic approximation is improved to take into account the complex ionized bubble distribution, (semi-)numerical simulations of patchy reionization are required to properly model the $\LyA$ opacity in the diffuse neutral IGM\footnote{One obvious improvement of the analytic model would be to introduce an outer radius $R_2$ for the first diffuse neutral patch, and construct a PDF for $R_2$ which can then be included into equation~(\ref{eq:analytic_bub}) to give 
\begin{equation}
e^{-\tau_{bub}^{\rm{eff}}(\nu_e) }\approx\int P(R_1)dR_1 \int e^{-\tau_{patch}(\nu_e,R_1,R_2)}P(R_2|R_1),
\end{equation} 
where $P(R_2|R_1)$ denotes the conditional probability of $R_2$, given $R_1$. We have started to include such improvement in our model. However, due to the difficulty in finding an analytic fitting function for $P(R_2|R_1)$, we have deferred this to a future work.}.


\subsection{The Red Damping Wing in Web Models}\label{sec:LyA_RT_web}

The ensemble of small-scale absorbers can also form a damping wing feature in the effective optical depth towards $\LyA$ galaxies as shown in the bottom panel of Fig.~\ref{fig:dampingwing_eff_HIpatch}, where the blue lines and the shaded areas refer to the mean transmission $\exp(-\tau_{web}^{\rm{eff}})$ and to the $1\sigma$ dispersion $\sigma_{\tau_{web}}^2=\langle (e^{-\tau_{web}}-e^{-\tau_{web}^{\rm{eff}}})^2\rangle$ as a function of $\Delta v$. 

Similarly to the bubble model, Fig.~\ref{fig:dampingwing_eff_HIpatch} indicates that the damping wing in web models becomes more opaque with increasing neutral fraction and decreasing $\Delta v$. 
Neutral fractions $\fHI_V\sim10^{-2}$ (W2 and W3), i.e. much higher than the one in the post-reionized universe (which is $\fHI_V\sim 10^{-4}$), are required to produce a $\sim60-80\%$ reduction of $\LyA$ visibility at $\Delta v=600~\rm km~s^{-1}$. On the other hand, the effective optical depth in W1 (which has a neutral fraction closer to $\sim 10^{-4}$) is $e^{-\tau_{\rm eff}}>0.9$ at $\Delta v=600~\rm km~s^{-1}$, i.e. it hardly affects the $\LyA$ visibility\footnote{This justifies the calibration of the intrinsic model discussed in \S~\ref{sec:mock}. However, the left bottom panel of Fig.~\ref{fig:dampingwing_eff_HIpatch} shows that $\exp(-\tau_{web}^{\rm{eff}})\sim 0.6$ at $\Delta v\sim 100~\rm km~s^{-1}$ even with a neutral fraction as small as the one in the post-reionized universe. For galaxies that have $\Delta v<200~\rm km~s^{-1}$ the impact of small-scale absorbers at $z<6$ should therefore be taken into account.}.
The scatter around the effective optical depth is again large, with $\sigma_{\tau_{web}}\sim0.2$. 

Finally, a comparison between the effective optical depth in the web and bubble models (e.g. B2 vs. W3 in Fig.~\ref{fig:dampingwing_eff_HIpatch}) shows that small-scale absorbers can produce a profile and scatter of the red damping wing similar to those of the bubble models. This explains the similarity in the $\LyA$ LFs observed through the large-scale bubbles and small-scale absorbers. 

\subsubsection{Comparison to the Analytic Model}

\begin{figure}
\centering
  \includegraphics[angle=0,width=\columnwidth]{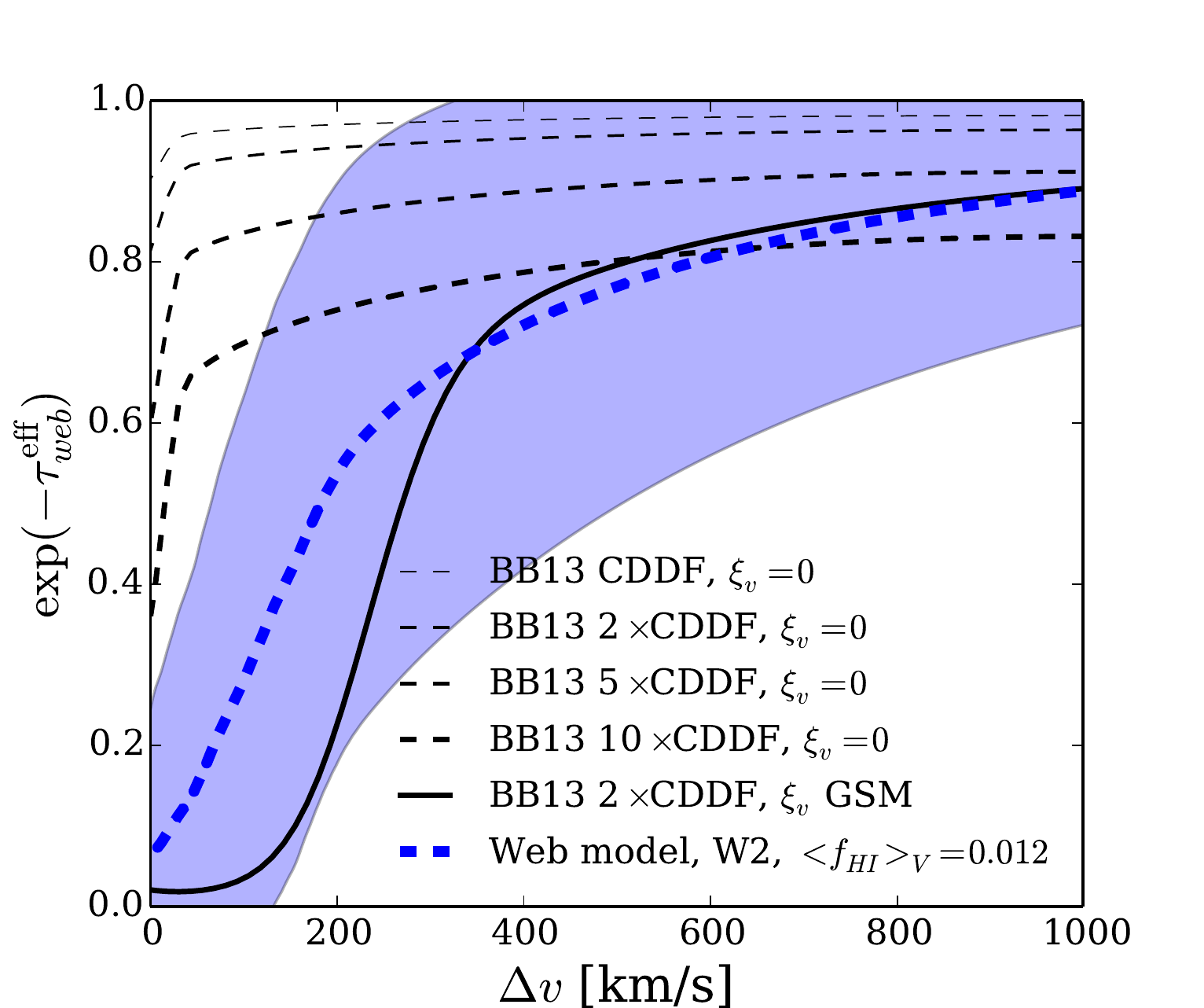}
  \caption{Comparison between the analytic and simulated web model. The black dashed lines are the effective optical depth from the analytic approximation (equation~(\ref{eqtauweb}) using the BB13 CDDF without the velocity-space galaxy-absorber correlation function, i.e. $\xi_v=0$. The black solid line instead includes a Gaussian streaming model (GSM) for $\xi_v$. This figure demonstrates the importance of the velocity-space galaxy-absorber correlation function.}
  \label{fig:tau_eff_compare}
\end{figure}

Fig.~\ref{fig:tau_eff_compare} compares the simulation and the analytic effective optical depth described by equation (\ref{eqtauweb}) in \S~\ref{LyA_RT_webtopo}. The black dashed lines refer to the analytic model without the effect of clustering and velocity field, i.e. $\xi_v=0$ in equation~(\ref{eqtauweb}), while the black solid line uses the Gaussian streaming model for $\xi_v$, i.e. equation~(\ref{xi_Gaussian}). The analytic model employs a factor of 2-10 boost to the extrapolated CDDF fit of \cite{2013MNRAS.436.1023B} at $z\simeq7$ (hereafter BB13 CDDF) to mimic the rapidly increasing abundance of small-scale absorbers. Our fiducial value is $2\times\rm CDDF$ (see \S~\ref{sec:CDDF} a discussion on the reason of this choice). 

Fig.~\ref{fig:tau_eff_compare} shows clearly that we cannot reproduce the results from our simulation by {\it only} changing the CDDF amplitude, while the agreement is much better if we simultaneously change the CDDF amplitude {\it and} include the galaxy-absorber correlation function in velocity-space (see \S~\ref{sec:xi} for the reason of the discrepancy at $\Delta v<200\rm~km/s$). In other words, both the abundance of small-scale absorbers and their velocity-space clustering around galaxies play a key role in determining the $\LyA$ visibility. In the following sections, we discuss in more detail the impact of ({\it i}) changing the CDDF (\S~\ref{sec:CDDF}) and ({\it ii}) the galaxy-absorber clustering (\S~\ref{sec:xi}).

\subsubsection{CDDF and $N_{\rm HI}$-Dependence of the Optical Depth}\label{sec:CDDF}
\begin{figure}
\centering
\advance\leftskip+0.1cm
  \includegraphics[angle=0,width=\columnwidth]{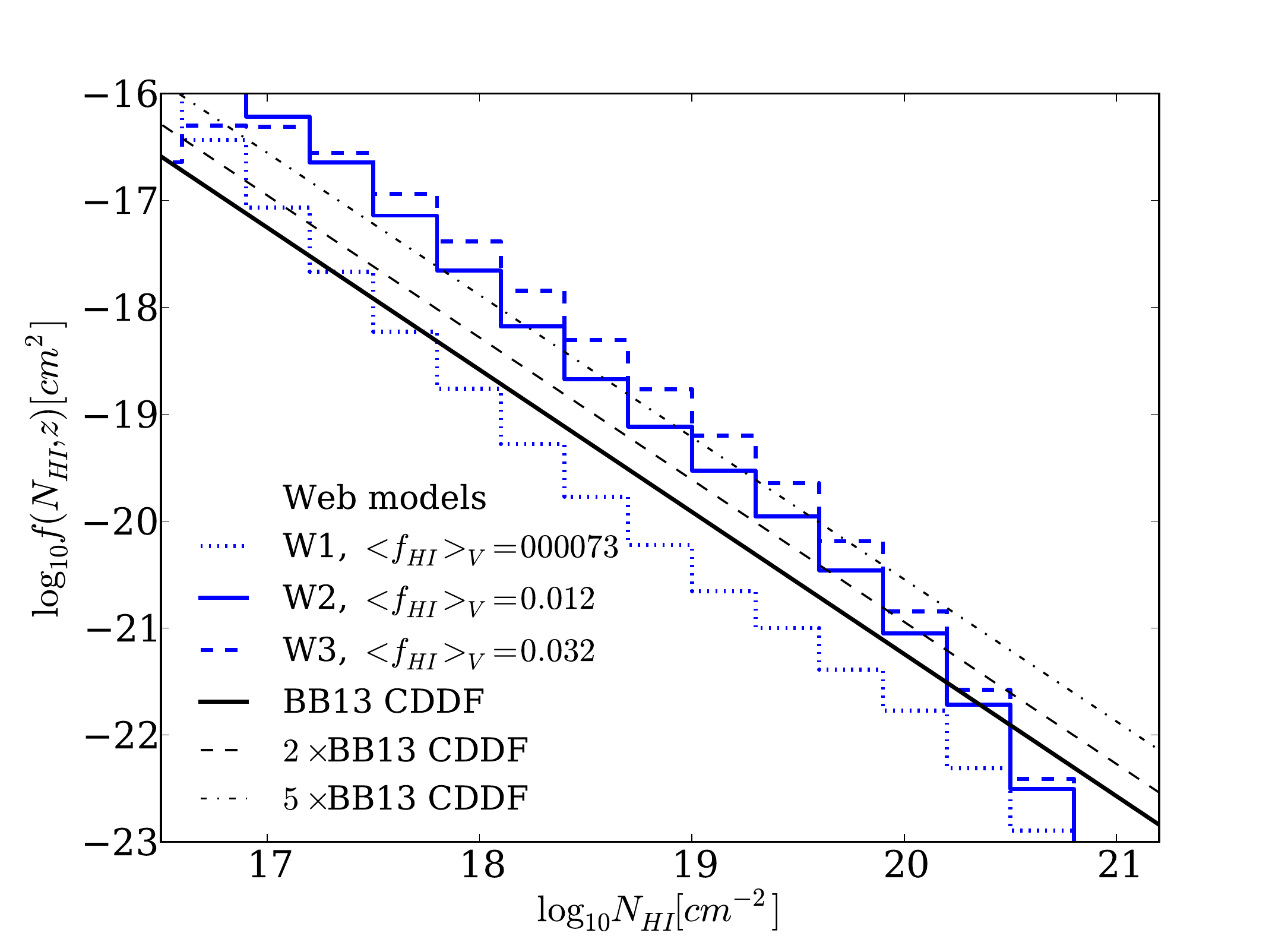}  
  \caption{$\HI$ Column Density Distribution Functions (CDDFs) extracted from web model simulation W1 (blue dotted line), W2 (blue solid), and W3 (blue dashed). The black lines refer to the extrapolation to $z\simeq7$ of the Becker \& Bolton (2013) fit (BB13 CDDF; solid), multiplied by a factor of 2 (dashed) and 5 (dotted).}\label{CDDF_simulated}
\end{figure}

We first justify the artificial boosting factor of the power-law CDDF adopted in the analytic model. Fig.~$\ref{CDDF_simulated}$ compares the CDDF obtained in our web model simulations\footnote{We have computed the CDDF by taking the projected column density over 10 cells. The highest $\NHI$ bins ($\log_{10}\NHI/{\rm cm^{-2}}\sim21.3$ and $20.6$) are about $\sim0.3$ dex larger than those calculated with a single cell, but the numbers converge for larger projected lengths. The effect is minor in the other bins.} 
to the BB13 CDDF with a factor of 1, 2 and 5 boost. The adopted boosts broadly mimic the increase in simulated CDDF amplitude due to lower photoionization rate/higher neutral fraction ($\Gamma=10^{-14}$~s$^{-1}$ and $5\times10^{-15}$~s$^{-1}$ for W2 and W3), although the slope is not properly reproduced. The fiducial choice of 2 (corresponding to $\Gamma\sim10^{-14}~\rm s^{-1}$) approximately represents the CDDF amplitude in the range $10^{19}{\rm cm^{-2}}<\NHI<10^{20.7}{\rm cm^{-2}}$, which gives the highest contribution to the red damping opacity. 

\begin{figure}
\centering
  \includegraphics[angle=0,width=0.5\textwidth]{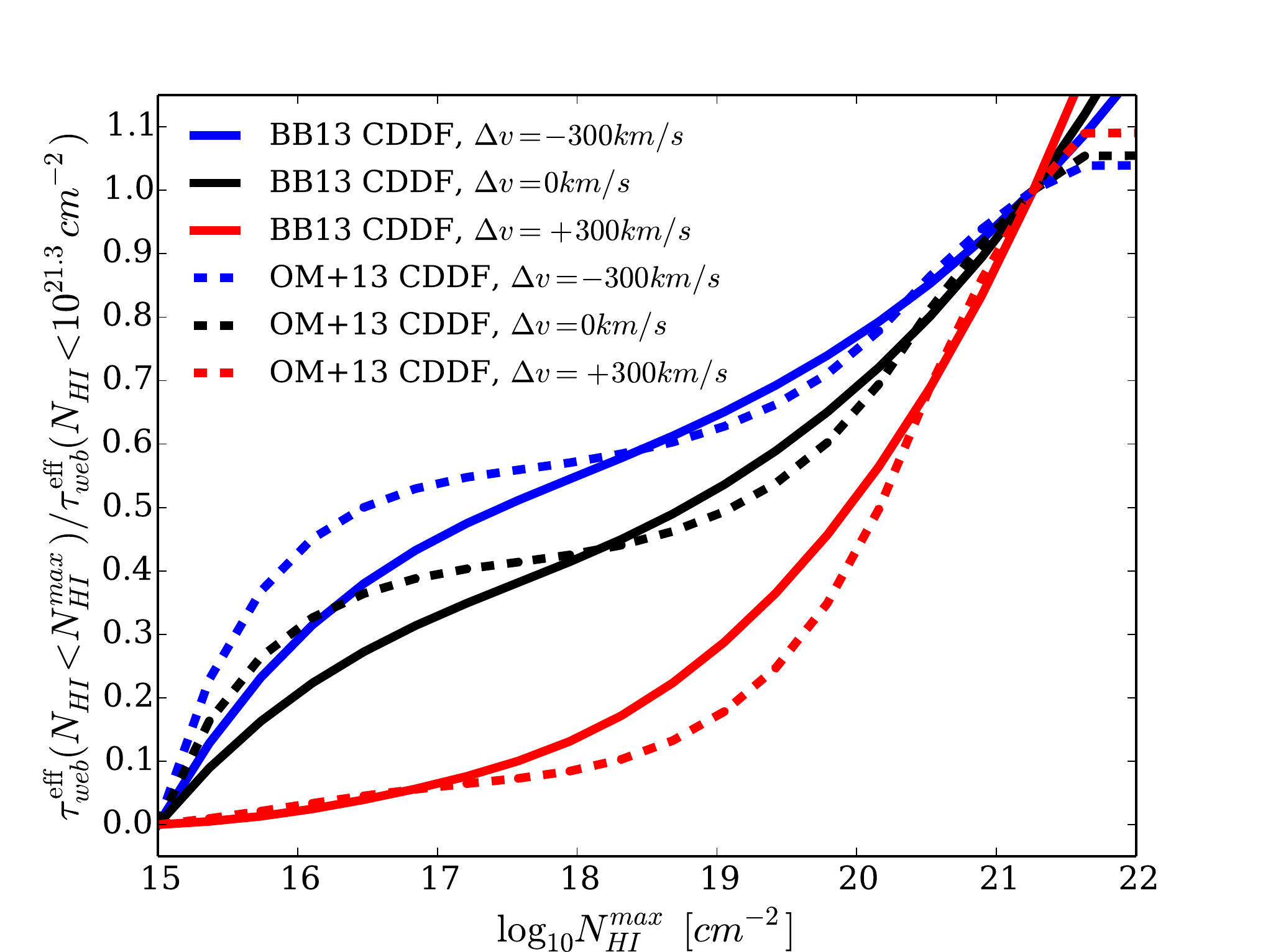}
  \caption{Cumulative contribution to $\tau_{\rm web}^{\rm eff}$ as a function of the maximum cut off column density $\NHI^{max}$.  Three different colours correspond to the optical depth redward ($\Delta v=300$~km/s, red line), at line centre ($\Delta v=0$~km/s, black line), and blueward ($\Delta v=-300$~km/s, blue line). The two functional forms of $\HI$ CDDF by Becker \& Bolton (2013) (solid lines) and O'Meara et. al. (2013) (dashed lines) are plotted. This figure shows that the optical depth redward of line centre, i.e. red damping wing, is mostly sensitive to strong $\HI$ absorbers with $\NHI>10^{19}~\rm cm^{-2}$, whereas the optical depth at smaller $\Delta v$ is increasingly more affected by weaker $\HI$ absorbers with $\NHI<10^{17-19}~\rm cm^{-2}$.}
   \label{fig:comulative_contribution}
\end{figure}

This predominance can be clearly seen in Fig.~\ref{fig:comulative_contribution}, which shows the ratio between the analytic effective optical depth from absorbers with column density below $\NHI^{max}$ and below $\NHI^{max}=10^{21.3}\rm cm^{-2}$, in a case with\footnote{The inclusion of the velocity-space correlation function, for example inflowing low column density absorbers, would enhance the contribution of lower column density absorbers to the optical depth.} $\xi_v=0$. 
 More than 80 per cent of the optical depth redward of line centre ($\Delta v=300$~km/s) comes from absorbers with $\NHI>10^{19}\rm cm^{-2}$, because of their prominent damping wings.  On the other hand, at $\Delta v=- 300$~km/s (i.e. blueward of the line resonance) lower column density absorbers with $\NHI<10^{18}\rm cm^{-2}$ can contribute $\sim50$ per cent to $\tau^{\rm eff}_{web}$ via resonant absorption. 
 
This strong dependence of the optical depth on the column density of absorbers is insensitive to the assumption about the shape of the CDDF, as shown by a comparison between the solid and dashed lines in Fig \ref{fig:comulative_contribution}, which refer to models using a CDDF from \cite{2013MNRAS.436.1023B} and \cite{2013ApJ...765..137O}, respectively. In both cases $\NHI\gtrsim10^{19}\rm cm^{-2}$ absorbers dominate the red damping wing, with a difference of only $\sim10$ per cent. 

Hence, the red damping wing opacity mainly depends on the abundance of strong $\HI$ absorbers, e.g. super-LLSs and DLAs, around $\LyA$-emitting galaxies. Their rapid increase (stronger than what expected from a simple extrapolation to $z>6$ of lower-$z$ CDDF) provides a large red damping wing opacity. 

\begin{figure}
\centering
\advance\leftskip-0.1cm
  \includegraphics[angle=0,width=\columnwidth]{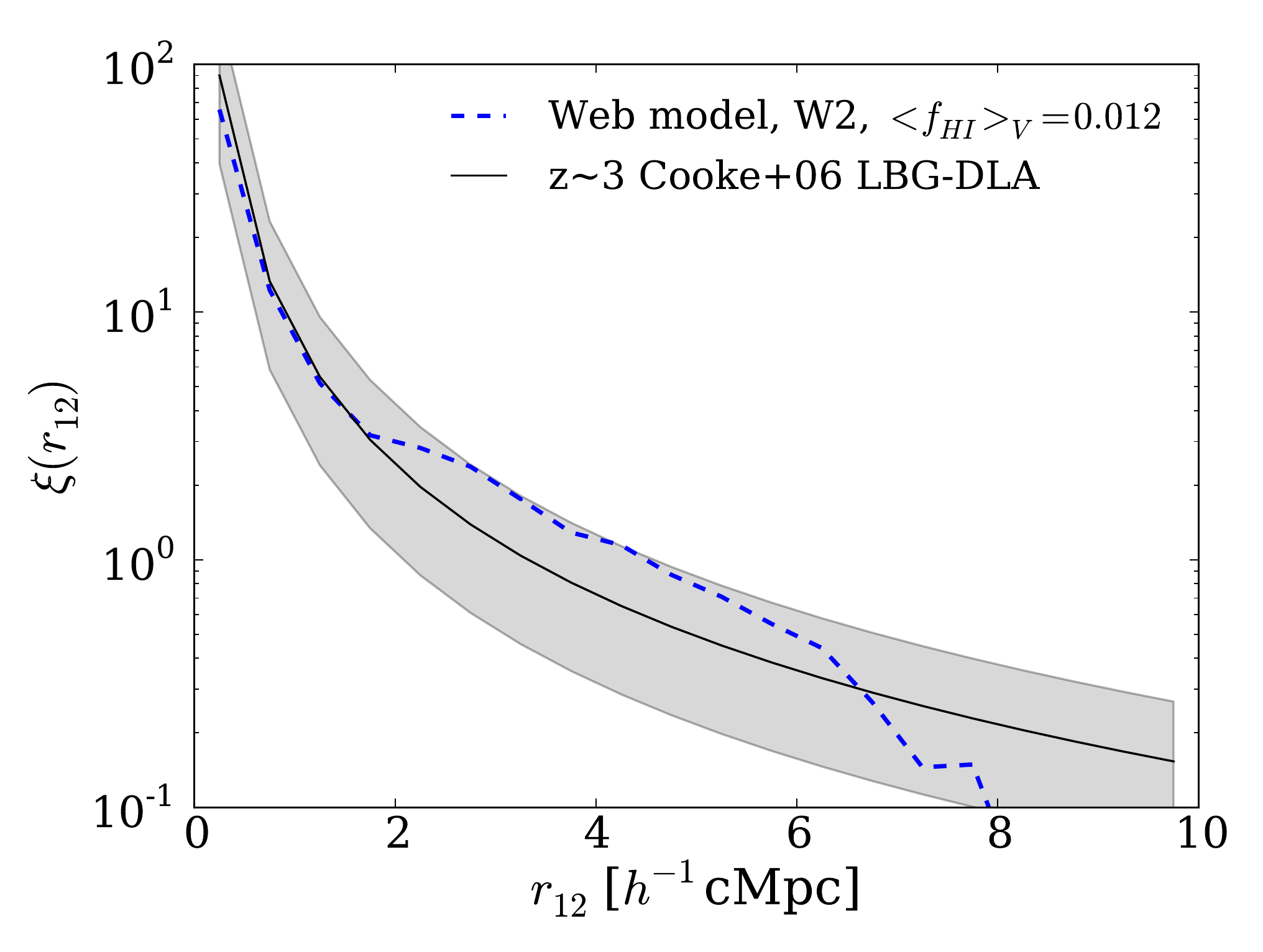}
  \caption{Simulated real-space galaxy-absorber correlation function at $z=7$ (dashed blue line) and best-fit to the LBG-DLA cross-correlation function observed by Cooke et al. (2006b) at $z\sim3$ (solid black). The latter is $\xi(r_{12})=(r_{12}/r_0)^{-\gamma}$, where $r_0=3.32\pm1.25h^{-1}\rm{cMpc}$, and $\gamma=1.74\pm0.36$. The shaded region spans the upper and lower $1\sigma$ errors in the observed correlation length for a fixed slope $\gamma=1.74$. The figure demonstrates the clustering of small-scale absorbers around galaxies.}
   \label{fig:2PCF}
\end{figure}

\begin{figure}
\centering
  \includegraphics[angle=0,width=\columnwidth]{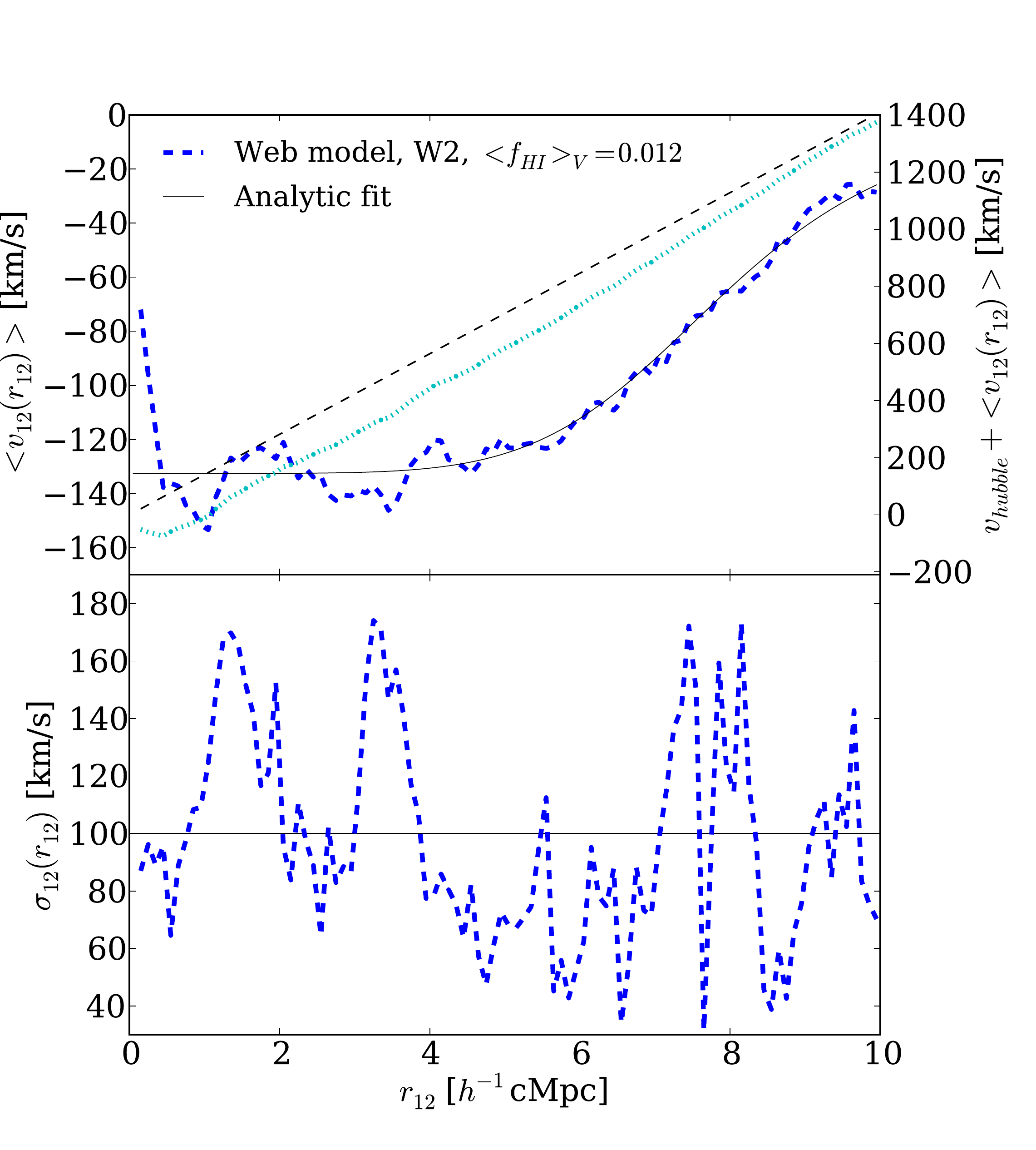}
  \caption{Galaxy-absorber mean pairwise velocity ({\it top panel}) and pairwise velocity dispersion ({\it bottom panel}) at redshift $z=7$. The blue line (with left y-axis) shows the proper mean pairwise peculiar velocity $\langle v_{12}(r_{12})\rangle$ and pairwise velocity dispersion $\sigma_{12}(r_{12})$ between absorbers and galaxies in simulation W2, while the solid black lines are the fits adopted in the analytic calculation with the Gaussian streaming model. The cyan line corresponding to the right y-axis of the top panel refers to the total proper pairwise velocity, $v_{hubble}+\langle v_{12}(r_{12})\rangle$, and the black dashed line is the Hubble flow velocity $v_{hubble}=H(z_s)r/(1+z_s)$. This figure shows the presence of cosmological inflow of absorbers onto galaxies, which slows down the total outflowing velocity including the Hubble flow.}
  \label{fig:v12}
\end{figure}

\subsubsection{Galaxy-Absorber Correlation Function in Velocity-Space}\label{sec:xi}

The galaxy-absorber correlation function in velocity space, $\xi_v$, is another key factor in the formation of the red damping wing. In the Gaussian streaming model of equation (\ref{xi_Gaussian}), $\xi_v$ depends both on ({\it i}) the real-space correlation function, $\xi(r_{12})$, and ({\it ii}) the galaxy-absorber pairwise mean velocity field $\langle v_{12}(r_{12})\rangle$, and pairwise velocity dispersion $\sigma_{12}(r_{12})$. 

The simulated real-space galaxy-absorber correlation function at $z=7$ is shown in Fig.~\ref{fig:2PCF}, together with the LBG-DLA correlation function observed by \cite{2006ApJ...636L...9C,2006ApJ...652..994C} at $z\sim 3$. The simulated $\xi(r_{12})$ is obtained by correlating the position of the galaxies and of the cells with $\xHI>0.9$ (which represent for us self-shielded absorbers) using the $\xi(r_{12})=DD/RR-1$ estimator (\citealt{1983ApJ...267..465D}). Clustering of self-shielding gas in the vicinity of $\LyA$-emitting galaxies is clearly important, and the simulated real-space correlation function appears (maybe surprisingly) similar to its lower-redshift observed counterpart.
We thus adopt the \cite{2006ApJ...652..994C} correlation function for our Gaussian streaming model in Fig.~\ref{fig:tau_eff_compare}. 

The mean pairwise velocity between $\LyA$-emitting galaxies and absorbers defined above is shown in the top panel of Fig. \ref{fig:v12}, both  in terms of the proper peculiar velocity $\langle v_{12}(r_{12})\rangle$ (blue lines) and of the total proper velocity $H(z_s)r_{12}/(1+z_s)+\langle v_{12}(r_{12})\rangle$ (cyan lines). The solid black line is the best-fit curve to the mean pairwise velocity, $\langle v_{12}(r_{12})\rangle=-v_{in}/[1+(r_{12}/r_v)^{\gamma_v}]$ where $v_{in}=133$~$\rm km/s$, $r_v=6.3h^{-1}\rm cMpc$ and $\gamma_v=6.2$. This is adopted to evaluate the Gaussian streaming model in Fig.~\ref{fig:tau_eff_compare}. For simplicity, rather than using a fit to the curve, we assume  a constant pairwise velocity dispersion equal to its mean, i.e. $\sigma_{12}=100~\rm km/s$.

\begin{figure*}
\centering
\advance\leftskip-2cm
  \includegraphics[angle=0,width=1.25\textwidth]{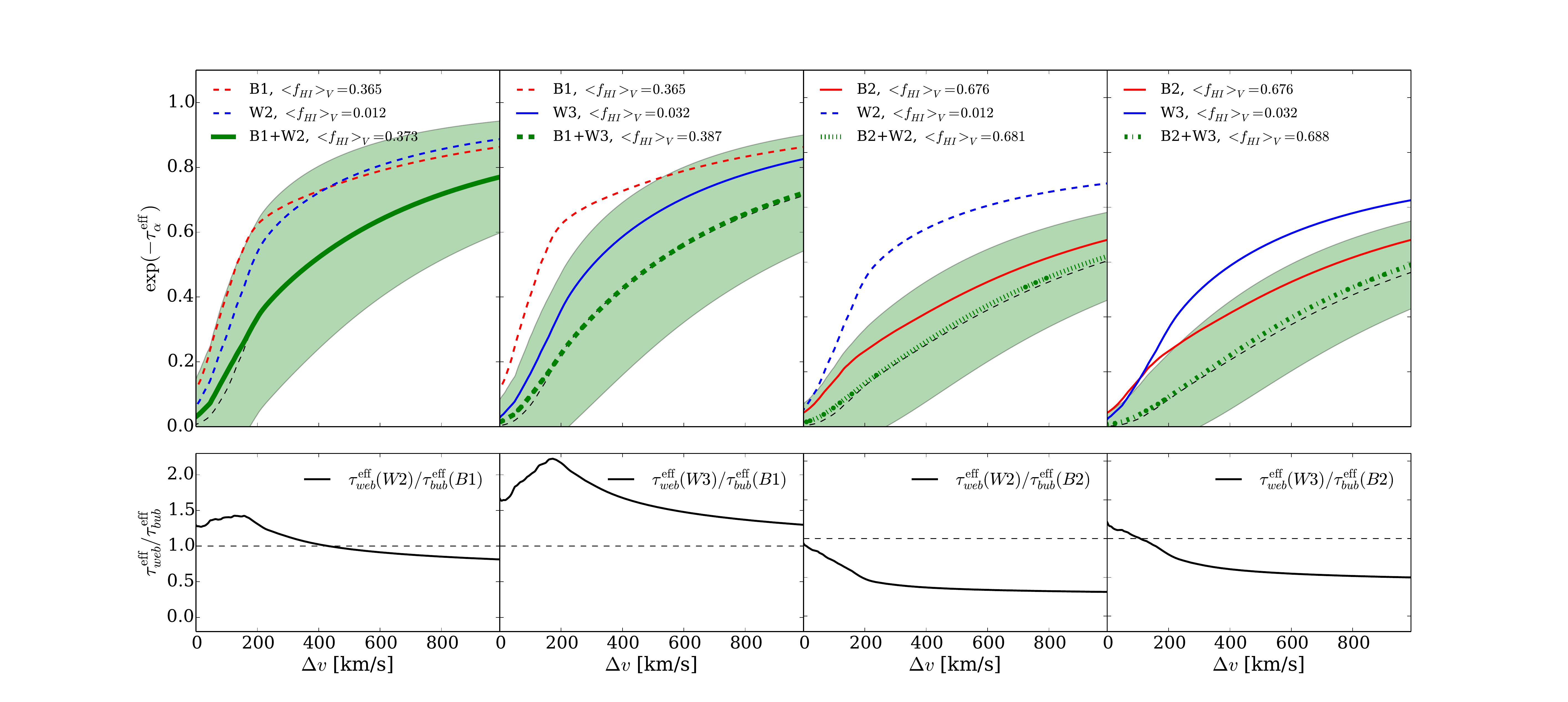}
  \caption{({\it Top panels}) $\LyA$ effective optical depth extracted from the hybrid web-bubble model simulations (green lines; B1+W2, B1+W3, B2+W2, and B2+W3 from left to right), together with the corresponding $1\sigma$ dispersion (shaded regions). The red and blue lines are the optical depth extracted from the bubble and web models used to construct the web-bubble models, and the black lines show the sum of these two contributions, i.e. $\tau^{\rm eff}_{bub}+\tau^{\rm eff}_{web}$. ({\it Bottom panels}) 
Ratio of effective optical depth between web and bubble models used for the corresponding web-bubble models. This shows the impact of large-scale bubbles and small-scale absorbers on the total optical depth as a function of $\Delta v$.
}
   \label{fig:web_bubble_ratio}
\end{figure*}

As shown in Fig.~\ref{fig:tau_eff_compare}, the impact of the galaxy-absorber correlation function in velocity-space provides an additional boost of effective optical depth relative to the model with $\xi_v=0$.  In fact, the enhanced clustering of absorbers around galaxies (Fig.~\ref{fig:2PCF}) renders the IGM more opaque. Furthermore, the cosmological inflow of absorbers onto galaxies (Fig.~\ref{fig:v12}) causes a departure from the Hubble flow in the immediate surroundings of galaxies and enhances the velocity-space clustering (the slower the total outflow velocity in the proper unit is, the more opaque to $\LyA$ photons the gas becomes, as it is less redshifted out of resonance). This can increase $\tau_{web}^{\rm{eff}}$, preferentially at the lower $\Delta v$. Thus, the effective optical depth including a velocity-space galaxy-absorber clustering is larger and steeper than the one including only a change in the CDDF amplitude (with $\xi_v=0$).

\vspace{1cm}
Overall, Fig.~\ref{fig:tau_eff_compare} shows that the simulation and the analytic model agree at $\Delta v>400$~$\rm km/s$, while our analytic approximation overestimates the opacity at $\Delta v<300$~$\rm km/s$, probably because we assume that the same galaxy-absorber correlation function applies to the full column density range of absorbers. This may lead to low column density absorbers with a $\xi(r_{12})$ which is too large. To address this issue, it is necessary to investigate in more detail the column density dependent clustering, the pairwise velocity field with outflow, and/or the effect of photoionization from the central galaxy.

\subsection{$\LyA$ Red Damping Wing in Web-Bubble Models}\label{sec:taueffwebbubble}

The top panels of Fig.~$\ref{fig:web_bubble_ratio}$ show the effective optical depth in the hybrid web-bubble models directly calculated from the simulations, together with the 1$\sigma$ dispersion of optical depth among different lines-of-sight. Not surprisingly, the red damping wing becomes more opaque towards higher neutral fractions, and the scatter from sight-line to sight-line is large. 
The red and blue lines show the contributions to the total simulated optical depth from the bubble and web models used to build each web-bubble model, while the black lines indicate the optical depth that we obtain by simply adding the two contributions, i.e. $\tau_{bub}^{\rm eff}+\tau^{\rm eff}_{web}$.  
A comparison between the optical depth extracted from the web-bubble simulations and a sum of the optical depths extracted from the corresponding bubble and web models indicates that the two agree very well\footnote{A slight discrepancy arises because the simple sum counts twice the neutral gas outside ionized bubbles (in the form of $\HI$ patches in bubble models and small-scale absorbers in web models), while in the simulations small-scale absorbers are present only within ionized bubbles by construction. The simple sum is thus expected to result in a slightly higher optical depth.}.

The bottom panels of Fig.~$\ref{fig:web_bubble_ratio}$ show the relative contribution of small-scale absorbers and large-scale $\HI$ patches to the total damping wing opacity. We find that:
\begin{itemize}
\item In all our web-bubble models, neither component dominates the total optical depth, as $\tau_{web}^{\rm{eff}}/\tau_{bub}^{\rm{eff}}\sim0.5-1.5$. 
\item The relative contribution depends on the intrinsic $\LyA$ line shift. The small-scale absorbers contribution increases with decreasing $\Delta v$ because their opacity is enhanced by the inflow onto galaxies (see \S~\ref{sec:xi}). On the other hand, the $\HI$ patches are typically located at a distance $\sim5-10h^{-1}$cMpc from $\LyA$-emitting galaxies (see Fig.~\ref{fig:bubble_size_PDF}), where the Hubble flow already dominates the total velocity. Therefore, they are more prominent at larger $\Delta v$. 
\end{itemize}
The above two points underline the importance of correctly modelling small-scale absorbers within the large-scale bubble morphology. This section concludes our discussion on the average impact of large-scale neutral patches (\S~\ref{sec:LyA_RT_bub}) and small-scale absorbers (\S~\ref{sec:LyA_RT_web}) on the $\LyA$ red damping wing opacity in a unified web-bubble framework. 

\begin{figure}
\centering
  \includegraphics[angle=0,width=0.5\textwidth]{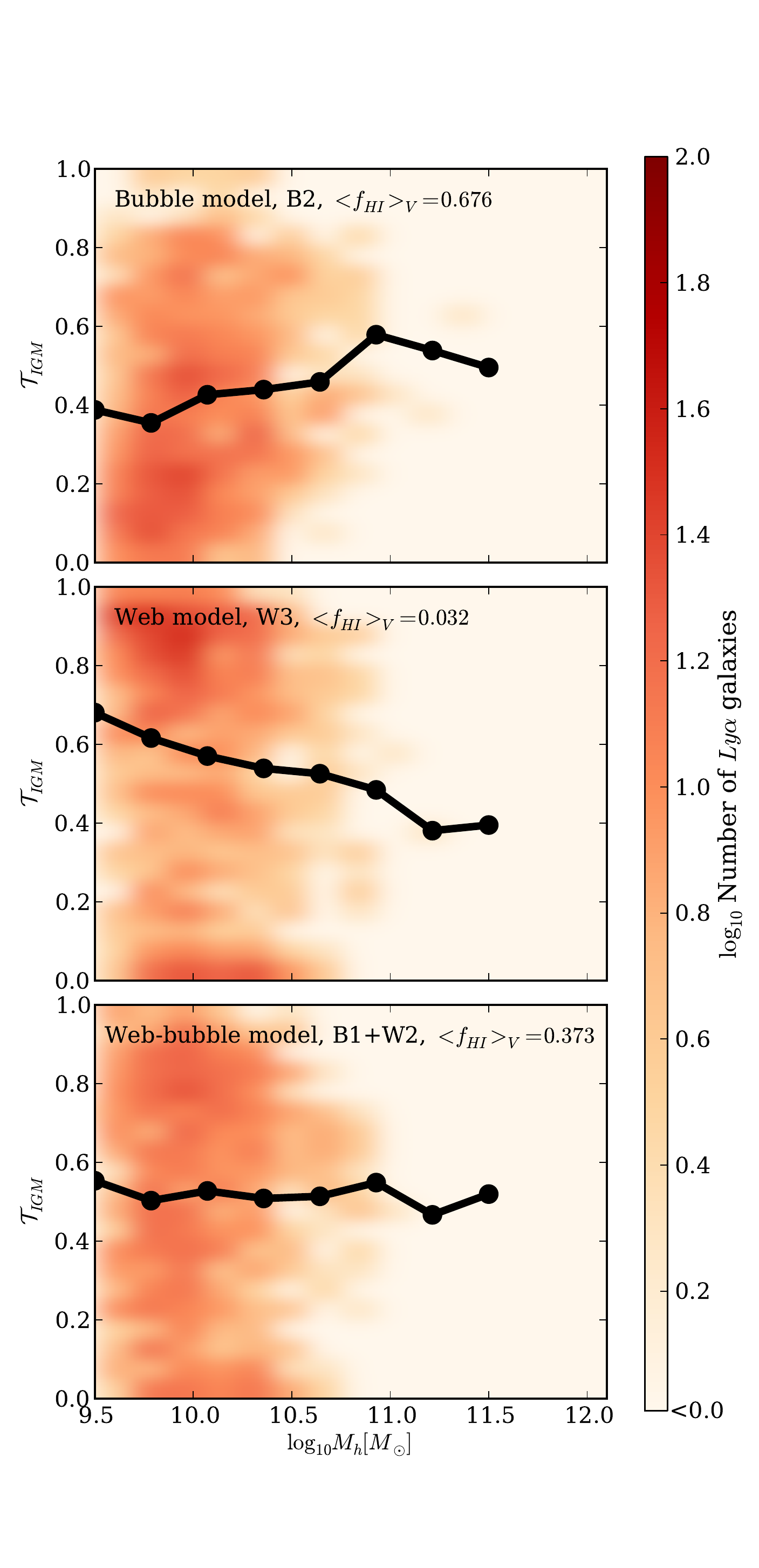}
  \caption{IGM transmission factor $\mathcal{T}_{IGM}$ along the line-of-sight to a $\LyA$-emitting galaxy as a function of the host halo mass $M_h$. The panels refer to the bubble model B2 (top), the web model W3 (middle), and the web-bubble model B1+W2 (bottom). The three models have a similar $\LyA$ LF. The black lines are the average IGM transmission factors $\langle\mathcal{T}_{IGM}\rangle$ in each halo mass bin. The colour indicates the number of $\LyA$-emitting galaxies at each location of the $M_h-\mathcal{T}_{IGM}$ map, which is a $20({\rm log})\times20({\rm linear})$ grid in the range $10^{9}~{\rm M_\odot}\le M_h\le 10^{12.5}~{\rm M_\odot}$ and $0\le\mathcal{T}_{IGM}\le1$. The figure shows that the $\mathcal{T}_{IGM}$-PDF is unimodal for a bubble model and bimodal for a web model.}\label{fig:Trans_haloMass}
\end{figure}

\subsection{Probability Distribution Functions for $\mathcal{T}_{\rm IGM}$ }\label{sec:T_IGM_Mh}

Fig.~$\ref{fig:Trans_haloMass}$ shows the distribution of the IGM transmission factor, $\mathcal{T}_{IGM}$, along the line-of-sight to a $\LyA$-emitting galaxy as a function of the host halo mass for models B2, W3, and B1+W2. These models have been chosen because they have a similar LF (see Fig.~\ref{fig:LyAlumfunc}) and effective optical depth (see Figs.~\ref{fig:dampingwing_eff_HIpatch} and \ref{fig:web_bubble_ratio}), and therefore a similar average $\LyA$ visibility. The black lines are the average IGM transmission factor $\langle\mathcal{T}_{IGM}(M_h)\rangle$ for each halo mass bin. 

In the bubble model plotted in the top panel of Fig.~$\ref{fig:Trans_haloMass}$, $\langle\mathcal{T}_{\rm IGM}(M_h)\rangle$ increases with $M_h$, as massive [small] haloes typically reside in large [small] ionized bubbles (in the highest mass bins the trend is reversed because of the poor statistics). At the same time, there exists a population of lower mass haloes clustered around the more massive ones, which is therefore also embedded within large ionized bubbles. This explains the large scatter exhibit by $\mathcal{T}_{\rm IGM}$ for low halo masses. Furthermore, in bubble models sight-lines to most (if not all) galaxies pass through $\HI$ patches, meaning that the intrinsic luminosity of most galaxies is reduced, and explaining the {\it unimodality} of the $\mathcal{T}_{IGM}$ distribution (something that was pointed out previously by \citealt{2014MNRAS.444.2114J} and \citealt{2015MNRAS.446..566M}).

As in web models self-shielding absorbers cluster around the more massive haloes (\S~\ref{sec:xi}), $\langle\mathcal{T}_{\rm IGM}(M_h)\rangle$ decreases with increasing $M_h$. The still present large scatter in the distribution now appears to be {\it bimodal}, with a peak at $\mathcal{T}_{IGM}\sim1$ and a second one at $\mathcal{T}_{IGM}\sim0$. These peaks correspond to cases in which a line-of-sight intersects an absorber or not. Differently from what happens in the bubble model where the intrinsic luminosity of all galaxies is reduced, here a suppression is [is not] present depending on whether a small-scale absorber is [is not] aligned with a galaxy, hence the bimodality. Our results are consistent with those by \cite{2015MNRAS.446..566M}, who also find that a bimodal distribution is a characteristic of the attenuation by small-scale absorbers. 

In hybrid web-bubble models the IGM transmission factor is a product of large-scale bubbles and small-scale absorbers. Because of the different mass-dependence of $\mathcal{T}_{\rm IGM}$ in the two models, the total IGM transmission factor here depends only weakly on $M_h$, and no clear unimodality or bimodality in the distribution is visible. For example, the sight-lines present in the web model with $\mathcal{T}_{IGM}\sim 1$ are now more opaque due to the absorption from the $\HI$ patches between large-scale bubbles. 

It is therefore clear that the conditional $\mathcal{T}_{IGM}$-PDF at a given halo mass, $P(\mathcal{T}_{IGM}|M_h)$, or in short the $\mathcal{T}_{IGM}-M_h$ relation, differs for web, bubble and web-bubble models. In the next section, we search for observational signatures of this variation in the intergalactic environment.

\subsection{Simultaneously Constraining the $\HI$ Fraction and the Topology of Reionization}\label{sec:constraining}

We now examine the prospect of observationally constrain the global $\HI$ fraction and the topology of reionization simultaneously by combining various statistics of $\LyA$ emitting galaxies. 

\subsubsection{The Equivalent Width Distribution}

\begin{figure}
\centering
  \includegraphics[angle=0,width=\columnwidth]{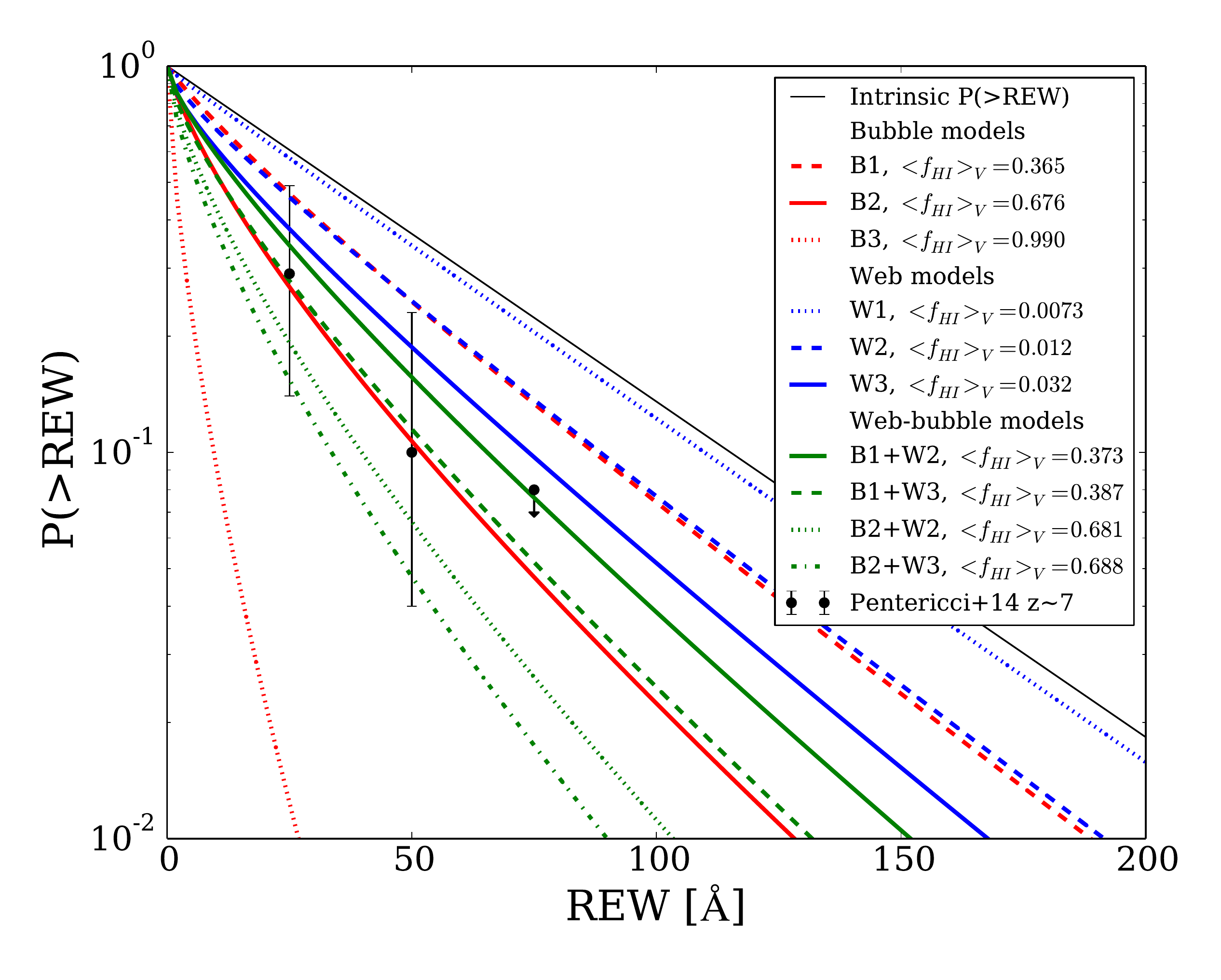}
  \caption{Cumulative probability distribution of the rest-frame equivalent width at $z=7$. The black line is the intrinsic REW distribution and the coloured lines refer to the observed REW distributions predicted from simulations: bubble model B1 (red dashed), B2 (red solid) and B3 (red dotted); web model W1 (blue dotted), W2 (blue dashed) and W3 (blue solid); web-bubble model B1+W2 (green solid), B1+W3 (green dashed), B2+W2 (green dotted) and B2+W3 (dotted-dashed). The black circles are the observation of Pentericci et al. (2014) without interloper correction (if the interloper correlation is taken into account the data points can be higher by $\sim20\%$).} 
   \label{fig:REW}
\end{figure}

Fig.~\ref{fig:REW} shows the cumulative probability distribution of the rest-frame equivalent width (REW), following the method of \cite{2011MNRAS.414.2139D},  
\begin{equation} 
P(>{\rm REW})=\int_0^1P_{\rm intr}(>{\rm REW}/\mathcal{T}_{IGM})P(\mathcal{T}_{IGM})d\mathcal{T}_{IGM},\label{REW_model}
\end{equation}
where the intrinsic REW distribution is $P_{\rm intr}(>{\rm REW_{intr}})=\exp(-{\rm REW_{intr}/REW_c})$, with ${\rm REW_c}=50~\rm \AA$ (\citealt{2011MNRAS.414.2139D}) and ${\rm REW_{intr}}={\rm REW}/\mathcal{T}_{IGM}$. The probability distribution function of the IGM transmission factor, $P(\mathcal{T}_{IGM})\propto\int P(\mathcal{T}_{IGM}|M_h)\frac{dn(>M_h)}{dM_h}dM_h$, is constructed from the simulations.

In all models, the observed REW distribution is decreased in comparison to the intrinsic one by an amount which increases with the $\HI$ fraction. Similarly to what observed for the Ly$\alpha$ luminosity function, a degeneracy is present between web and bubble models, with, for example, B1 and W2 providing similar REW distributions.

However, the degeneracy can be partially broken if the REW distribution is combined with the $\LyA$ LF. In fact, while models B2, W3 and B1+W2 are degenerate in $\LyA$ LF (see Fig.~\ref{fig:LyAlumfunc}) they produce distinguishable observed REW distributions. Although this is not always the case (for example, B1 and W2 show similar curves both in the $\LyA$ LF and the REW distribution), such a combined analysis offers a test to differentiate reionization models.

The argument above can be better understood by noting that the observed $\LyA$ LF and REW distribution depend differently on the $\mathcal{T}_{IGM}-M_h$ relation. To see this, we first express the $\LyA$ LF in terms of $P(\mathcal{T}_{IGM}|M_h)$ as
\begin{equation}
\frac{dn(>L^{obs}_\alpha)}{dL_\alpha}=\int P(L^{obs}_\alpha|M_h)\frac{dn(>M_h)}{dM_h}dM_h,
\end{equation}
where
\begin{align} 
&P(L_\alpha^{obs}|M_h)= \nonumber \\
&\int_0^1P_{\rm intr}(L_\alpha^{obs}/\mathcal{T}_{IGM}|M_h)P(\mathcal{T}_{IGM}|M_h)d\mathcal{T}_{IGM}.\label{CLF}
\end{align}
$P_{\rm intr}({L_\alpha}|M_h)$ is the intrinsic conditional probability distribution of the $\LyA$ luminosity given a halo mass\footnote{Explicitly, we use $P_{\rm intr}(L_\alpha|M_h)=\delta_D[L_\alpha-L_\alpha(M_h)]$ as we assume a one-to-one mapping between $L_\alpha$ and $M_h$ based on the abundance matching technique.}. A comparison between equations (\ref{REW_model}) and (\ref{CLF}) shows a different dependence on $P(\mathcal{T}_{IGM}|M_h)$\footnote{Note that equation (\ref{REW_model}) implicitly assumes that the intrinsic REW distribution is independent of halo mass. We can, of course, generalize this modelling to include the halo mass dependence, but because this in general differs from the one of the $\LyA$ luminosity, the dependence of the two statistics on the $\mathcal{T}_{IGM}-M_h$ relation is expected to differ as well.}. This is because the Ly$\alpha$ LF is constructed from Ly$\alpha$ selected LAEs, while the REW-PDF is constructed from continuum selected galaxies. In fact, \citet{DW12} and \citet{Gronke15} have shown that selection by Ly$\alpha$ line flux enhances the contribution of UV-faint galaxies (at fixed Ly$\alpha$ flux), which are absent from continuum selected samples. As such UV-faint galaxies should preferentially reside in low mass haloes, this difference in selection function would introduce a different dependence in the $\mathcal{T}_{IGM}-M_h$ relation that may lead to a drop in the observed $\LyA$ LF different from the one in the REW distribution.  

Hence, a combined analysis of $\LyA$ LF and REW distribution may allow to constrain the $\HI$ fraction and the topology of reionization. 
We can already do this analysis. The upper limit at $\rm REW=75~\AA$ slightly favours the bubble or web-bubble models with $\fHI_V\sim68\%$ or $\sim37\%$. If we include this constraint, the neutral fraction is favored to be of order of tens of per cent. This constraint is very weak because of a large uncertainty due to the interloper contamination. Moreover, the same observations favour bimodal quenching of the $\LyA$ visibility, which is associated with web-models. This argument simply illustrates that a combined analysis of Lyα LF and REW-PDF can shed light on the history and topology of reionization.

\subsubsection{$\LyA$ Fraction of Lyman-Break Galaxies}\label{sec:chi_lya}

\begin{figure}
\centering
  \includegraphics[angle=0,width=0.5\textwidth]{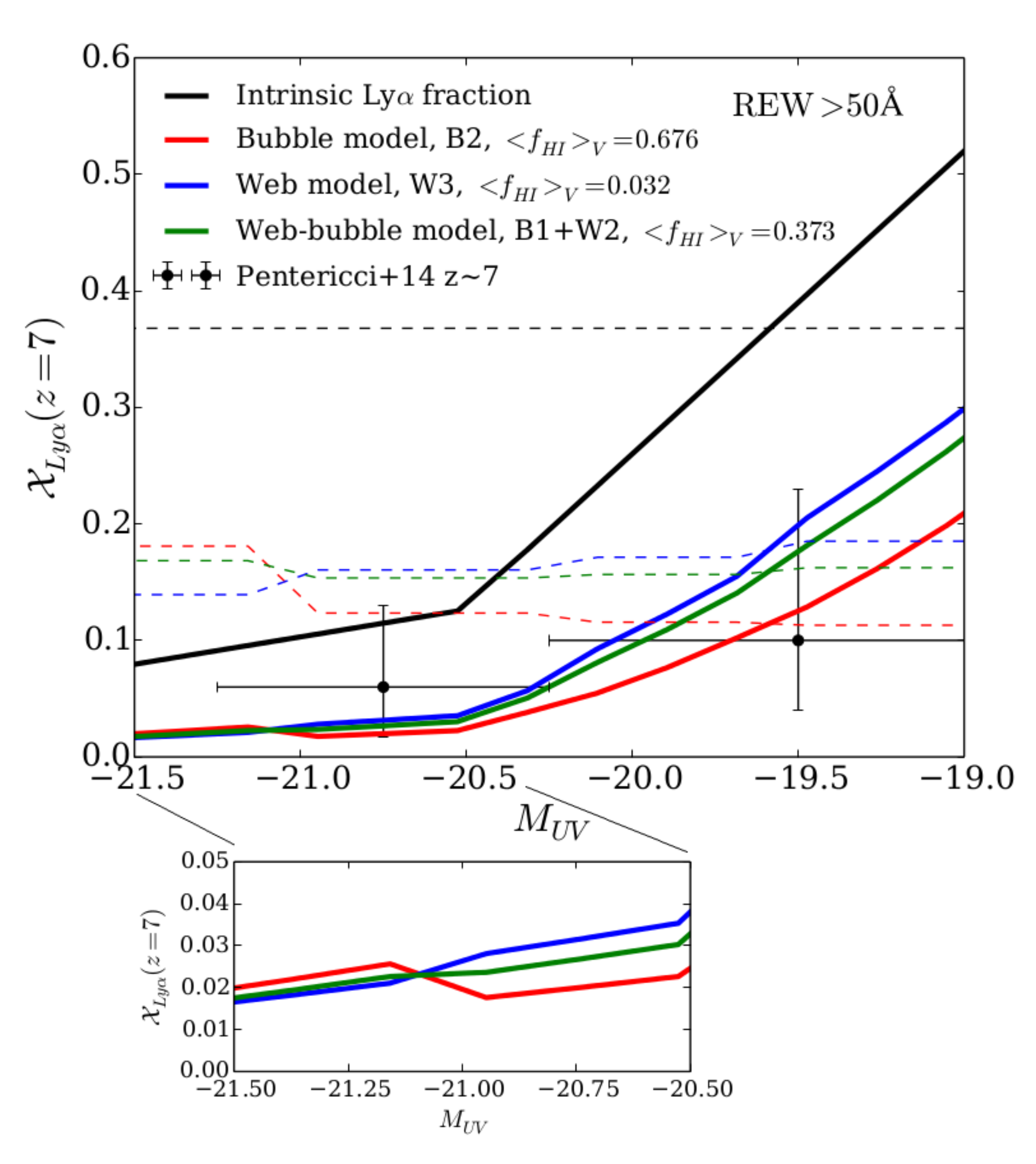}
\caption{$\LyA$ fraction of LBGs having $\rm REW>50~\AA$ at $z=7$ as a function of the UV magnitude $M_{UV}$ of galaxies. The line style indicates the intrinsic fraction obtained with the $M_{UV}$-dependent (solid lines) and the uncorrelated (dashed lines) model (see text for more details), and the line colours refer to the intrinsic $\LyA$ fraction (black), and to the observed $\LyA$ fraction for models B2 (red), W3 (blue) and B1+W2 (green). The three models have a similar $\LyA$ LF (Fig.~\ref{fig:LyAlumfunc}). The black points are the observations of Pentericci et al. (2014), where the horizontal error bars indicate the bin size used. The smaller plot is a zoom-in of the $\LyA$ fraction - $M_{UV}$ relation to emphasize the upturn at UV-bright LBGs caused in model B2 by their larger $\LyA$ visibility.}
   \label{fig:lyafraction}
\end{figure}

The power of such joint analysis can be strengthened once the $M_{UV}$-dependent $\LyA$ fraction of LBGs measurement is included as well.
The $\LyA$ fraction of LBGs (hereafter $\mathcal{X}_{\mbox{\tiny Ly$\alpha$}}$) is defined as the fraction of LBGs with a UV magnitude $M_{UV}$ and $\LyA$ REW greater than a given value.
We generalize the method of \cite{2011MNRAS.414.2139D} (see also \citealt{DW12}) to calculate the $\LyA$ fraction of LBGs as
\begin{align} 
&\mathcal{X}_{\mbox{\tiny Ly$\alpha$}}(>{\rm REW}|M_{UV})= \nonumber \\
&\int_0^1\mathcal{X}_{\mbox{\tiny Ly$\alpha$}}^{\rm intr}(>{\rm REW}/\mathcal{T}_{IGM}|M_{UV})P(\mathcal{T}_{IGM}|M_{UV})d\mathcal{T}_{IGM},
\end{align}
where $\mathcal{X}_{\mbox{\tiny Ly$\alpha$}}^{\rm intr}(>{\rm REW_{intr}}|M_{UV})$=$e^{-{\rm REW_{intr}}/{\rm REW_c}(M_{UV})}$ is the intrinsic $\LyA$ fraction, ${\rm REW_c}(M_{UV})$ is a characteristic REW (see Appendix \ref{app:LyAfrac} for more details), and the conditional $\mathcal{T}_{IGM}$ probability distribution function at a given $M_{UV}$ is
\begin{equation}
P(\mathcal{T}_{IGM}|M_{UV})=\int P(\mathcal{T}_{IGM}|M_h)P(M_h|M_{UV}) dM_h .
\end{equation}
 We construct our intrinsic model assuming that UV-bright LBGs populate more massive haloes, and consider a case with a correlation between REW and $M_{UV}$ ($M_{UV}$-dependent model) and one with no correlation (uncorrelated model). The $M_{UV}$-dependent model is our fiducial case because observations suggest that such correlation exists (\citealt{2010MNRAS.408.1628S}; \citealt{2013ApJ...772...99J}, but see \citealt{2009MNRAS.400..232N}). More details are provided in Appendix \ref{app:LyAfrac}.

Fig.~\ref{fig:lyafraction} shows the intrinsic and observed $\LyA$ fractions at $z=7$ for models B2, W3, and B1+W2, which all produce a similar $\LyA$ LF. Two main features emerge:
\begin{itemize}   
\item the bubble model shows an {\it upturn} of $\LyA$ fraction at UV-bright LBGs (typically defined as those with $M_{UV}<-20.25$), 
while the web model shows a monotonic decrease of $\LyA$ fraction for increasing UV-bright LBGs\footnote{Note that the downturn in the bubble model B2 at $M_{UV}\lesssim -21.15$ is due to the poor statistics, similarly to what observed in the top panel of Fig.~\ref{fig:Trans_haloMass}.}. This qualitative change in the shape of the $M_{UV}$-dependent $\LyA$ fraction is robust against different intrinsic models of REW. 
\item In the $M_{UV}$-dependent model (solid lines), the drop in the observed $\LyA$ fraction compared to the intrinsic one is larger for UV-faint LBGs ($M_{UV}>-20.25$) than for UV-bright LBGs in {\it all} models. The common expectation that the drop of $\LyA$ fraction of UV-faint LBGs is larger than the one of UV-bright LBGs occurs {\it only} in bubble models (\citealt{2012ApJ...744...83O}) is true {\it only if} the intrinsic ${\rm REW}$ and $M_{UV}$ are uncorrelated (dashed lines).
\end{itemize}
 
The upturn of the $M_{UV}$-dependent $\LyA$ fraction can be understood as an imprint of the $\mathcal{T}_{IGM}-M_h$ relation (see \S~\ref{sec:T_IGM_Mh}). In fact, because UV-bright LBGs in bubble models are more likely to be surrounded by large ionized bubbles, the probability that their $\LyA$ emission is visible (i.e. that they are associated to larger $\mathcal{T}_{IGM}$) is higher than for UV-faint LBGs. On the other hand, in web models the small-scale absorbers cluster more strongly around UV-bright LBGs, lowering their $\LyA$ visibility. The upturn of $\LyA$ fraction, therefore, does not happen in web models. 
As a consequence, the qualitative change in the {\it shape} of the $M_{UV}$-dependent $\LyA$ fraction can be used as an indicator of the (possible) presence of large-scale bubbles. 

On the other hand, a drop of the $\LyA$ fraction for UV-faint LBGs larger than for UV-bright LBGs  cannot be used as a decisive evidence of patchy reionization. In fact, while in the uncorrelated case (dashed lines) we indeed see a larger drop for UV-faint LBGs only for the bubble model, in the $M_{UV}$-dependent case (solid lines) such drop is visible for all models. The simplest explanation for this is that because, to first order approximation, the neutral IGM suppresses the $\LyA$ emission by re-scaling the characteristic REW as $\langle \mathcal{T}_{IGM}\rangle {\rm REW_c}(M_{UV})$ (see also Appendix~\ref{app:LyAfrac}), UV-faint galaxies (with an intrinsically larger $\rm REW_c$) experience a larger reduction in number above a given REW than the UV-bright galaxies (with intrinsically small $\rm REW_c$) do. 

In summary, the analysis of the $M_{UV}$-dependent $\LyA$ fraction of LBGs provides a powerful diagnostic tool to characterize the impact of large-scale bubbles and small-scale absorbers when properly interpreted. Hence, when combined with the $\LyA$ LF, it offers an opportunity to constrain the $\HI$ fraction and the topology of reionization simultaneously.
While the aim of the present paper is to highlight the potential of this diagnostics, we plan to use it more extensively in a future study.

\section{Discussion \& Conclusions}\label{sec:conclusion}

The visibility of $\LyA$-emitting galaxies during the Epoch of Reionization is controlled by both diffuse $\HI$ patches in the IGM, and small-scale self-shielding absorbers around galaxies. It is therefore important to correctly include small-scale absorbers inside large-scale ionized bubbles. In this work we have explored the impact of both large-scale bubbles and small-scale absorbers on the visibility of the population of $\LyA$-emitting galaxies at $z>6$, using a powerful combination of an analytic approach and hydrodynamical simulations, which covers the full range of models explored in recent investigations (\citealt{2013MNRAS.428.1366J}; \citealt{2013MNRAS.429.1695B}; \citealt{2015MNRAS.446..566M}; \citealt{2015MNRAS.452..261C}). We have considered the IGM $\LyA$ RT in three different classes of IGM ionization structure, namely ({\it i}) the bubble model, where only large-scale ionized bubbles due to patchy reionization are present, ({\it ii}) the web model, where only small-scale absorbers are considered, and ({\it iii}) the web-bubble model, which includes both small-scale absorbers and large-scale bubbles. 

Our main conclusions are:
\begin{itemize}
  \setlength\itemsep{1em}
\item  
The observed $\LyA$ LF evolution from $z=5.7$ to $z\sim7$ requires a neutral fraction $\fHI_V\sim60-80\%$ in bubble models, $\fHI_V\gtrsim 1-3\%$ in web models, and $\fHI_V\sim30-70\%$ in web-bubble models.  

\item A sole analysis of the $\LyA$ luminosity function or of the distribution of rest frame equivalent width cannot put a stringent constraint on the reionization history. The $\LyA$ LF function and the REW-PDF can be equally suppressed in bubble, web, and web-bubble models, yet with very different global $\HI$ fractions. Hence, there is a fundamental degeneracy between the ionization structure of the IGM and the global $\HI$ fraction inferred from $\LyA$ surveys (see \S~\ref{sec:Lya_LF}). 

\item We showed in \S~\ref{sec:constraining} that a joint analysis of the Ly$\alpha$ LF and the REW-PDF of LBGs can improve the constraints on the neutral fraction by breaking the degeneracy with the topology of reionization. 

\item The $\LyA$ fraction of LBGs can be a powerful diagnostic to study the relative importance of large-scale $\HI$ patches and small-scale absorbers in the IGM. We caution that a drop in $\LyA$ fraction that is larger for UV-faint LBGs than for UV-bright LBGs (as in \citealt{2012ApJ...744...83O}) can be reproduced with web and web-bubble models, and does not provide exclusive evidence for patchy reionization. Instead, we argue that the shape of the $M_{UV}$-dependent $\LyA$ fraction may provide more insight into the topology of reionization (see e.g. Fig.~\ref{fig:lyafraction}). 

For example, an upturn of $\LyA$ fraction for UV-bright LBGs can be caused by large-scale ionized bubbles, but also by an increase in the UV background around UV-bright galaxies, which reduces the abundance of small-scale absorbers. 
Interestingly, this upturn may already have been observed at $4.5<z<6$ (\citealt{2010MNRAS.408.1628S}), and may reflect large fluctuations in the UV background. These fluctuations have been proposed to explain observations of the cumulative effective optical depth distribution at $z\gtrsim5$ in the spectra of high-redshift QSOs (\citealt{2015MNRAS.447.3402B}; \citealt{2015arXiv150501853C}). 

\item Our analytic formalism shows that the $\LyA$ damping wing opacity from small-scale absorbers is highly influenced by the clustering and the pairwise velocity field of galaxy-absorber pairs (see  \S~\ref{sec:LyA_RT_web}). Absorbers with $\NHI>10^{19}\rm~cm^{-2}$, i.e. super-LLS/DLAs, provide the largest contribution to the the red damping wing at $\Delta v>300$ km s$^{-1}$, while lower column density absorbers are important at smaller $\Delta v$. Understanding the galaxy-absorber correlation functions and their velocity fields can improve the robustness with which the reionization history can be constrained using $\LyA$ emitting galaxies. Direct observational constraints on $\HI$ CDDF and galaxy-absorbers correlation function (and as a function of $N_{\rm HI}$) can therefore be very useful. A possible approach is to extend to the range $3<z<7$ the survey strategy that searches for $\LyA$-emitting galaxies in the foregrounds of high-redshift QSOs, similar to the observation of \cite{2006ApJ...652..994C}, Keck Baryonic Structure Survey (\citealt{2012ApJ...750...67R}; \citealt{2014MNRAS.445..794T}), and VLT LBG Redshift Survey (\citealt{2011MNRAS.414...28C}). This observational strategy is already within reach at $z\sim5.7$ (\citealt{2014MNRAS.442..946D}). 

\item We showed that the total effective optical depth in web-bubble models can be written as the sum of those in web and bubble models, i.e. $\tau_{\alpha}^{\rm eff}\approx\tau_{bub}^{\rm eff}+\tau^{\rm eff}_{web}$ (see \S~\ref{sec:taueffwebbubble}). This is an important result as fast semi-numeric simulations can be used to generate $\tau_{bub}^{\rm eff}$. These simulations can then be complemented with (improved) analytic or possibly empirical prescriptions for $\tau^{\rm eff}_{web}$ (as in \S~\ref{sec:LyA_RT_web}) to efficiently generate more realistic web-bubble models. 

\item Web, bubble and web-bubble models produce different $\mathcal{T}_{\rm IGM}$-PDFs (\S~\ref{sec:T_IGM_Mh}). Bubble models show a unimodal $\mathcal{T}_{\rm IGM}$-PDF, while small-scale self-shielding absorbers in the web-model have a bimodal $\mathcal{T}_{\rm IGM}$-PDF. The modality of the hybrid web-bubble model depends on which component dominates the IGM opacity. \cite{2014ApJ...793..113P} have provided observational evidence for bimodal quenching of $\LyA$ flux (see Treu et al. 2012, 2013 for details on the procedure). Our results imply that bimodal quenching indicates an influence of small-scale absorbers on the Ly$\alpha$ visibility (also see Mesinger et al. 2015), which is opposite to the common interpretation. 
\end{itemize}

In conclusion, in this paper we have shown that a joint analysis of different statistics of Ly$\alpha$ emitting galaxies (e.g. $\LyA$ LF, REW distribution, $\LyA$ fraction of LBGs, correlation function), can break degeneracies associated with individual probes. It should therefore be possible to constrain simultaneously the global $\HI$ fraction {\it and} the reionization topology, when armed with a suit of models of reionization in which both large-scale bubble morphology and small-scale absorbers are included.

\section{Acknowledgment}
K.K. thanks Hannes Jensen, Martin Haehnelt, Michele Sasdelli for useful comments and discussions, Andrew Chung for carefully reading the manuscript, Akira Konno and Masami Ouchi for kindly providing the data points shown in Fig.~\ref{fig:LyAlumfunc}, and Romain Teyssier and the RAMSES developer team to make the code public and user friendly.

\bibliographystyle{mn2e}
\bibliography{Reference}


\appendix
\section{The mass-weighted neutral fraction in the post-reionized universe}\label{A1}
The mass-weighted $\HI$ fraction in the post-reionized universe can be estimated from DLA/LLS surveys and $\LyA$ forest observations, which measure the $\HI$ column density distribution function. As follows, this quantity can then be
converted into the $\HI$ fraction embedded as $\LyA$ absorbers, 
such as DLA, LLS, and diffuse IGM.

\begin{figure}
\centering
  \includegraphics[angle=0,width=0.5\textwidth]{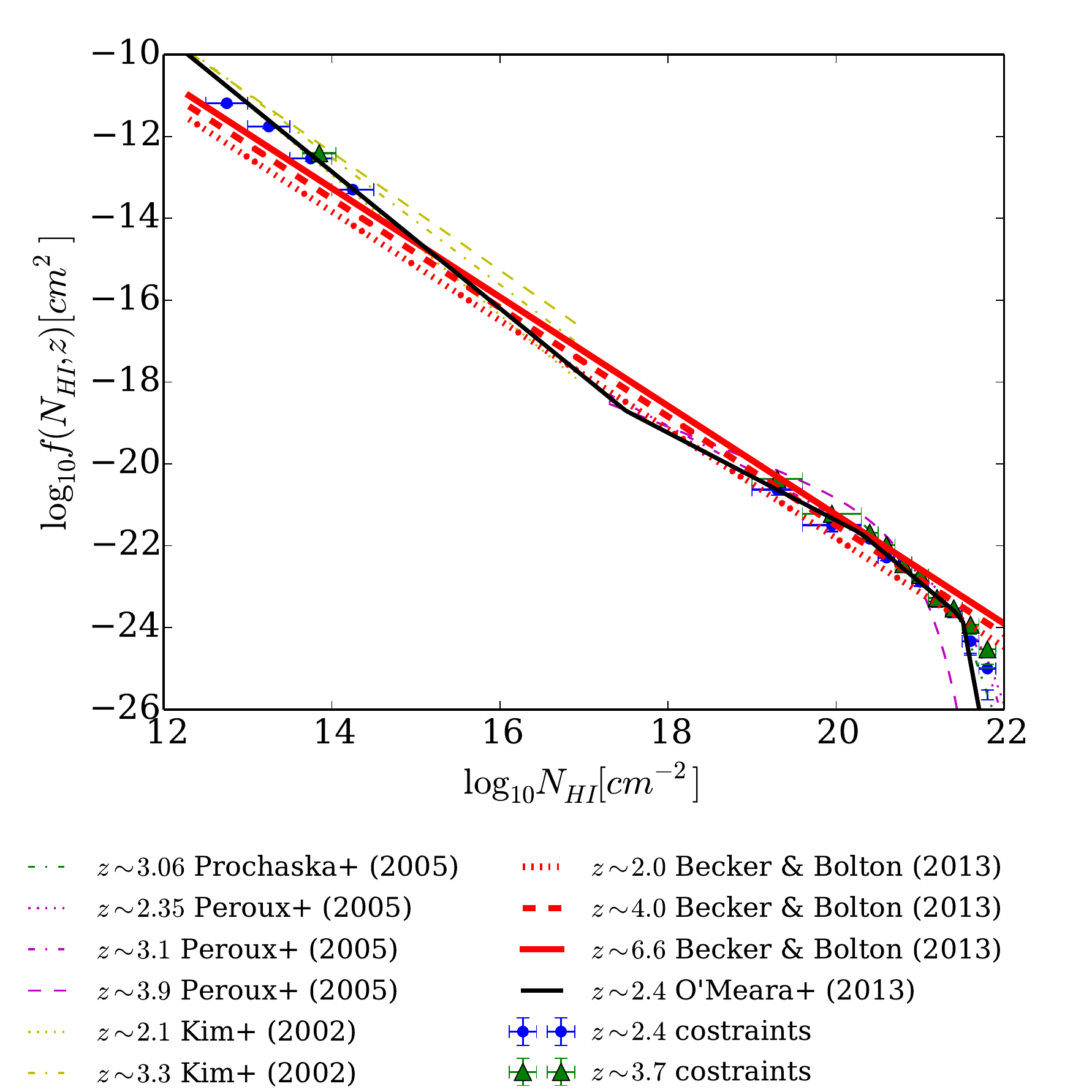}
  \caption{$\HI$ column density distribution function $f(\NHI,z)$ at $z\gtrsim2$. 
  The lines show the fits to the CDDF taken from the literature. The fit by Becker \& Bolton (2013) is used to extrapolate to $z>4$. The points at $z\sim2.4$ and $z\sim3.7$ use the compilation of data presented in O'Meara et al. (2013).}
   \label{fig:CDDF}
\end{figure}

The proper number density of $\HI$ gas in the universe, $\nHI^{prop}(z)$, is expressed as
(cf. \citealt{2009RvMP...81.1405M})
\begin{eqnarray}
\nHI^{prop}(z)&=&\int \NHI\frac{\partial^2 \mathcal{N}}{\partial\NHI\partial z}
\left|\frac{dz}{dl_p}\right|d\NHI, \\ \nonumber
                         &=&\frac{(1+z)^3H_0}{c}\int\NHI f(\NHI,z)d\NHI,
\end{eqnarray}
where $l_p$ is the proper distance, $dl_p/dz=c/H(z)(1+z)$.
Therefore, the fraction of neutral hydrogen over the total hydrogen atoms 
in the entire universe, $\fHI_M$,  
is given by $\fHI_M=\nHI^{prop}(z)/\bar{n}_{\rm{H}}^{prop}(z)$ 
\footnote{The fraction of total number of neutral hydrogen, $\mathcal{N}_\HI$, 
over the total hydrogen atom counts, $\mathcal{N}_{\rm H}$, 
is given by the mass-weighted neutral fraction 
$\fHI_M=\mathcal{N}_\HI/\mathcal{N}_{\rm{H}}=\int \xHI \nH dV/\int \nH dV=\int \xHI \rho dV/\int \rho dV$.
The volume-weighted and the mass-weighted neutral fraction are identical only for
a homogeneous IGM: $\fHI_M=\int \xHI \bar{\rho} dV/\int \bar{\rho} dV=\int \xHI dV/\int dV=\fHI_V$. 
}, 
\begin{equation}
\fHI_M=\frac{8\pi G m_H}{3H_0c(1-Y)\Omega_b}
\int_{\NHI^{min}}^{\NHI^{max}}\NHI f(\NHI,z)d\NHI, \label{HIfromCDDF}
\end{equation}
where $m_H$ is the mass of a hydrogen atom and 
$\bar{n}_H^{prop}(z)=\frac{3H_0^2(1-Y)\Omega_b}{8\pi G m_H}(1+z)^{3}
=2.057\times10^{-7}(1+z)^3\left(\frac{\Omega_b h^2}{0.023}\right)\rm cm^{-3}$ 
for a helium abundance $Y=0.25$.
The upper and lower limits of the integration specify whether the $\HI$ content is embedded
in the $\LyA$ forest absorbers ($\log_{10}[\NHI/\rm cm^{-2}]<17$), Lyman-limit systems ($17<\log_{10}[\NHI/\rm cm^{-2}]<20.3$),
or damped $\LyA$ systems ($20.3<\log_{10}[\NHI/\rm cm^{-2}]$). We integrate equation~($\ref{HIfromCDDF}$) using
the fitting functions to the observed CDDFs, $f(\NHI,z)$. We use the CDDF fitting functions from
\cite{2002MNRAS.335..555K} for the $\LyA$ forest absorbers, 
\cite{2003MNRAS.346.1103P} for the LLS range, and
\cite*{2005ApJ...635..123P} for the DLA range.
The observed $f(\NHI,z)$ and the various fits are shown in Fig.~\ref{fig:CDDF}. 

\section{Effective optical depth of dynamical small-scale absorbers}\label{app:tauweb}
The opacity from small-scale absorbers is determined by the phase-space distribution function of galaxy-absorber pairs, $f(r_{12},v_{12},\NHI)$, where $r_{12}$ is the comoving separation and $v_{12}$ is the peculiar pairwise radial velocity of pairs.

The line transfer is sensitive to the clustering in total velocity space, $v_c=aHr_{12}+v_{12}$. The probability to find an absorber within $v_c$ and $v_c+dv_c$ and column density $\NHI$ and $\NHI+d\NHI$ is $p(v_c,\NHI)dv_cd\NHI$. Then, the effective optical depth is given by (\citealt{1980ApJ...240..387P})
\begin{equation}
\tau_{web}^{\rm{eff}}=\iint p(v_c,\NHI)\left[1-e^{-\tau_{abs}(v_c,\NHI)}\right]dv_cd\NHI.
\end{equation}
$p(v_c,\NHI)$ is related to the phase-space distribution function of galaxy-absorber pairs through the transformation of variables $r_{12}, v_{12}$ to $v_c$,
\begin{align}
p(v_c,\NHI)&=\iint\delta_D\left[v_c-(aHr_{12}+v_{12})\right]f(r_{12},v_{12},\NHI)dv_{12}dr_{12}\nonumber \\
&=\int p_v(v_c-aHr_{12}|r_{12},\NHI)p_r(r_{12},\NHI)dr_{12},\label{B2}
\end{align}
where $\delta_D$ is the Dirac delta function.
For the second equality, we have used $f(r_{12},v_{12},\NHI)=p_v(v_{12}|r_{12},\NHI)p_r(r_{12},\NHI)$, where $p_v(v_{12}|r_{12},\NHI)dv_{12}$ is the conditional probability to find an absorber with peculiar pairwise velocity between $v_{12}$ and $v_{12}+dv_{12}$ at given pair separation $r_{12}$ and column density $\NHI$, and $p_r(r_{12},\NHI)dr_{12}d\NHI$ is the probability to find an absorber in the range $r_{12}$ to $r_{12}+dr_{12}$ and $\NHI$ to $\NHI+d\NHI$. The real-space correlation function $\xi(r_{12},\NHI)$ of absorbers around galaxies gives
\begin{equation}
p_r(r_{12},\NHI)=\frac{\partial^2\mathcal{N}}{\partial\NHI\partial z}\left|\frac{dz}{dr}\right|\left[1+\xi(r_{12},\NHI)\right],
\end{equation}
where $|dr/dz|=c/H(z_s)$. Substituting into equation (\ref{B2}), 
\begin{equation}
p_v(v_c,\NHI)=\frac{\partial^2\mathcal{N}}{\partial\NHI\partial z}\left|\frac{dz}{dr}\right|\frac{1}{aH}\left[1+\xi_v(v_c,\NHI)\right],
\end{equation}
where we have defined the absorber-galaxy correlation function in velocity space as
\begin{align}
&1+\xi_v(v_c,\NHI)\equiv \\
&~~aH\int dr_{12}\left[1+\xi(r_{12},\NHI)\right]p_v(v_c-aHr_{12}|r_{12},\NHI).
\nonumber
\end{align}
Thus, the effective optical depth is 
\begin{align}
\label{B6}
\tau_{web}^{\rm{eff}}&=\int d\NHI\frac{\partial^2\mathcal{N}}{\partial\NHI\partial z}\left|\frac{dz}{dr}\right|\times \\
&~~~~\int\frac{dv_c}{aH}(1+\xi_v(v_c,\NHI))\left[1-e^{-\tau_{abs}(v_c,\NHI)}\right].
\nonumber
\end{align}
All the quantities are evaluated at redshift $z=z_s$. By rearranging we obtain equation (\ref{eqtauweb}).

In the absence of clustering, $\xi_v=0$, the effective optical depth (\ref{B6}) reduces to the well-known expression for the Poisson-distributed absorbers $\tau_{web}^{\rm{eff}}=\int dz\int d\NHI \left|\frac{dl_p}{dz}\right|\frac{\partial^2\mathcal{N}}{\partial\NHI\partial l_p}(1-e^{-\tau_{abs}})$ (e.g. \citealt{1996ApJ...461...20H}). 

We show two examples of the velocity-space correlation function $\xi_v$. For a pure Hubble flow $v_c=aHr_{12}$, $p_v(v_{12}|r_{12},\NHI)=\delta_D(v_{12})$. Thus, $\xi_v(v_c)=\xi\left(r_{12}=\frac{v_c}{aH}\right)$. Furthermore,
a Gaussian streaming model is a simple generalization where the conditional pairwise peculiar velocity PDF is modelled as  $p_v(v_{12}|r_{12},\NHI)=\frac{1}{\sqrt{2\pi\sigma_{12}^2(r_{12})}}\exp\left[-\frac{(v_{12}-\langle v_{12}(r_{12})\rangle)^2}{2\sigma_{12}^2(r_{12})}\right]$, where $\langle v_{12}(r_{12})\rangle$ and $\sigma_{12}(r_{12})$ are the radial pairwise mean peculiar velocity and velocity dispersion, respectively.

\section{Abundance matching}\label{app:abundance_matching}

\begin{figure}
\centering
  \includegraphics[angle=-90,width=\columnwidth]{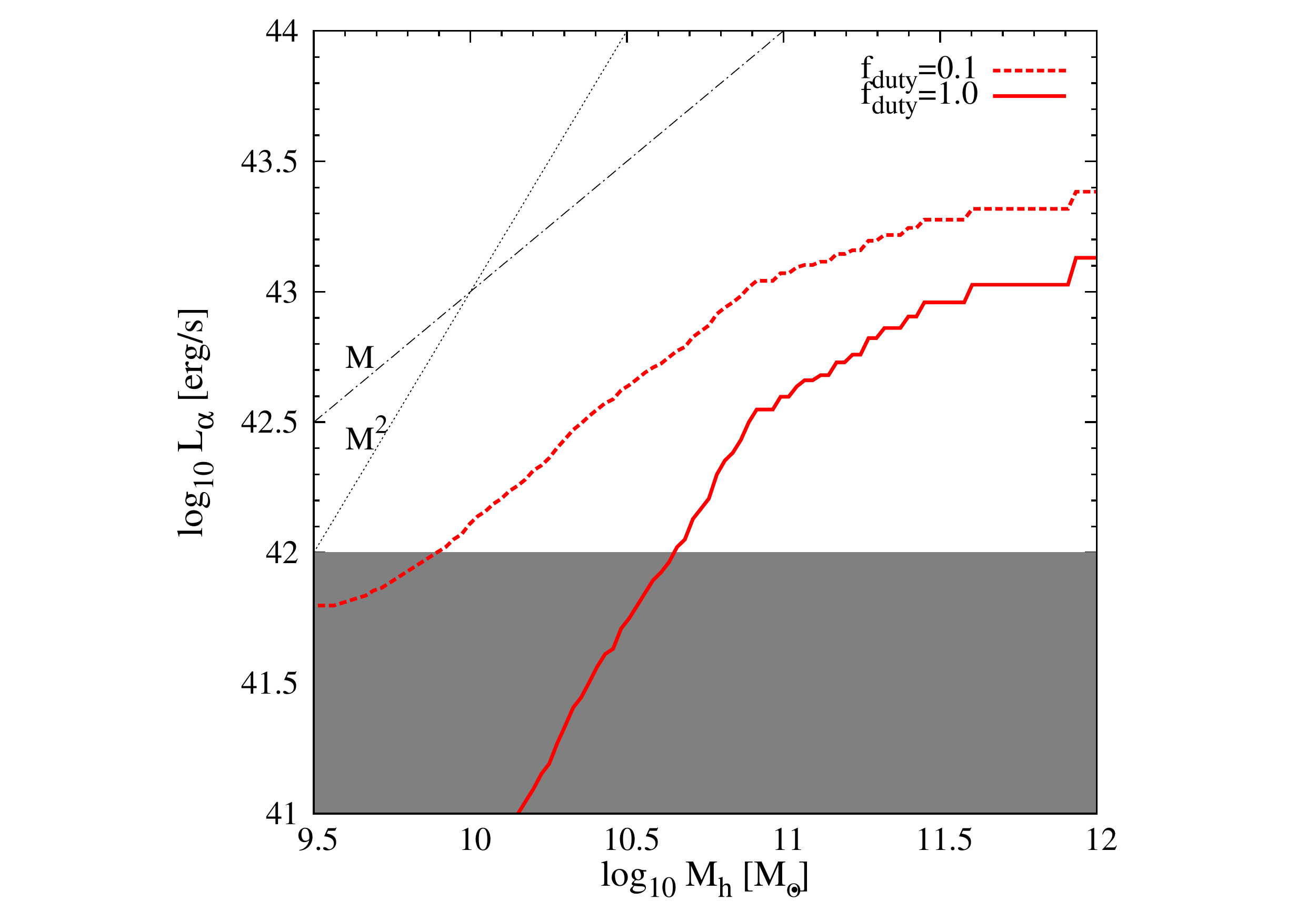}
  \caption{Semi-empirical relation between the halo mass and the intrinsic $\LyA$ luminosity from the abundance matching technique.
  The red solid (dashed) line is the result of matching the simulated halo mass function at $z=7$ with the $z=5.7$ observed $\LyA$ luminosity function for $f_{duty}=1.0~(0.1)$. The shadowed region indicates the luminosity range below detection limit. The two black dashed lines correspond to $L_\alpha\propto M_h,~M_h^2$.}
   \label{fig:abundance_matching}
\end{figure}

The abundance matching technique gives a semi-empirical relation between the halo mass and the $\LyA$ luminosity for each $f_{duty}$ as shown in Fig.~$\ref{fig:abundance_matching}$. The red lines are the result of matching the simulated halo mass function at $z=7$ with the observed $z=5.7$ $\LyA$ luminosity function (\citealt{2008ApJS..176..301O}) assuming a duty cycle $f_{duty}=0.1$ and 1.

Fig.~$\ref{fig:abundance_matching}$ shows that, given a halo mass, a higher duty cycle requires a brighter $\LyA$ luminosity to match the observed $z=5.7$ $\LyA$ luminosity function, and that a simple functional form, e.g. $L_\alpha\propto M_h,~M_h^2$, cannot match the semi-empirical relation.

In our model, the intrinsic $\LyA$ luminosity of each galaxy (halo) is assigned according to the $L_\alpha-M_h$ relation with $f_{duty}=1$ in Fig.~$\ref{fig:abundance_matching}$. 

\section{$\LyA$ RT through the IGM:
computing the line-of-sight skewers and optical depth}\label{appendix}

We compute the $\LyA$ optical depth in the red damping wing as follows. The density, temperature, velocity and local $\HI$ fraction fields along skewers originating at the location of halos and parallel to the z-axis are extracted from the hydrodynamical  and radiative transfer simulations. To obtain a converged numerical integration of the optical depth, the sampling size of the skewers, $\delta l$, must be sufficiently fine. To be on the safe side, the Doppler core of the Voigt line profile should be resolved. In the velocity space this is $\delta v/c=\Delta\nu_D/\nu_\alpha=4.286\times10^{-7}(T/{\rm K})^{1/2}$. Therefore, the velocity space resolution must be $\delta v\approx0.13(T/1{\rm K})^{1/2}$~km/s, which corresponds to a real space resolution of $\delta l=\delta v/H(z_s)\approx0.17(T/{\rm K})^{1/2}~\rm{pkpc}$ at $z_s=7$ with our cosmological parameters. If this criterion is not met, scattering by Doppler core could be missed. Although
the Doppler core scattering is important in low density regions to produce $\LyA$ forest absorption blueward of the rest-frame $\LyA$ line, here we are interested only in the red damping wing and the Lorentz wing scattering. Therefore, a converged evaluation of the optical depth in the red damping wing can still be obtained without strictly meeting this resolution criterion. 
Nonetheless, the sampling of the line-of-sight skewers must be sufficiently fine, and
a sub-sampling within a cell of the cosmological hydrodynamical simulations is required to obtain a convergence in equation ($\ref{numeric_optdpt}$).

To this aim, we have assumed that the density, ionization, temperature and peculiar velocity fields are constant within each cell, while the Hubble flow is allowed to vary. This is required to recover the analytic solution and to obtain a numerically converged optical depth in the limit of homogeneous expanding IGM. 

The discretized form of the optical depth is then integrated at each frequency point $\nu_e$ 
using the line-of-sight skewers  according to
\begin{equation}
\tau_\alpha(\nu_e)=\sum_{i=1}^N\sigma_\alpha \nHI(l_i)\varphi_\nu\left[T_i,\nu_e\left(1-\frac{v_{tot}(l_i)}{c}\right)\right]\delta l.
\label{numeric_optdpt}
\end{equation}
The maximum proper length of the line-of-sight skewers influences the far redward optical depth, as a lower length would results in more transmission. We choose the maximum proper length of the skewer to be $12$~pMpc. If a skewer exits the simulation box, a random cell in a random face of the box is chosen, and the line-of-sight is followed until the maximum proper length is reached.
We have verified that for a homogeneous expanding IGM, the result at $\Delta v\sim1000~\rm{km/s}$ has a discrepancy of $\sim8\%$ relative to the analytic solution of the optical depth. Because the IGM will become more ionized as $\LyA$ photons travel through the medium and because we retain the same redshift output to extract the line-of-sight skewers, we choose the maximum length of our skewer samples to be 12 pMpc. 

The lower bound of the optical depth integration is chosen to be $300h^{-1}\rm ckpc$. As a reference, the virial radius of a halo with mass $M_h$ is $R_{vir}\approx78.5(M_h/10^{11}h^{-1}M_\odot)^{1/3}h^{-1}\rm ckpc$, i.e. we exclude from the calculation the gas contained within a halo, as well as all the structures on scales  
smaller than the Jeans length because they are not well resolved in our simulations.

\section{Intrinsic $\LyA$ fraction}\label{app:LyAfrac}

We write the intrinsic $\LyA$ fraction as $\mathcal{X}_{\mbox{\tiny Ly$\alpha$}}^{\rm intr}(>{\rm REW_{intr}}|M_{UV})=e^{-{\rm REW_{intr}}/{\rm REW_c}(M_{UV})}$ where ${\rm REW_c}(M_{UV})$ is the characteristic REW. 

The $M_{UV}$-dependent model and uncorrelated model differ in their functional form of ${\rm REW_c}(M_{UV})$, as the latter assumes a constant ${\rm REW_c}(M_{UV})=50\rm~\AA$, while the former uses the ${\rm REW_c}(M_{UV})$ obtained from the best-fit to the $\LyA$ fraction of LBGs observed at $3<z<6$ (\citealt{2010MNRAS.408.1628S}), i.e. $\mathcal{X}_{\mbox{\tiny Ly$\alpha$}}^{\rm intr}(>{\rm REW}|M_{UV},z=7)=\mathcal{X}_{\mbox{\tiny Ly$\alpha$}}^{3<z<6}(>{\rm REW}|M_{UV})$. 

Furthermore, for $P(M_h|M_{UV})$ we assume a one-to-one mapping between UV magnitude and halo mass, i.e. $P(M_h|M_{UV})=\delta_D(M_h-M_h(M_{UV}))$. The $M_h-M_{UV}$ relation is given by $M_h(M_{UV})=M^\ast_h\times10^{-(M_{UV}-M^\ast_{UV})/2.5}$ where $M^\ast_h=10^{10}\rm~M_\odot$ and $M^\ast_{UV}=-19$. We note that this relation tends to assign masses which are typically lower than those derived from observations. For example,  $M_h(M_{UV}=-20)=2.5\times10^{10}~\rm M_{\odot}$, which is much lower than the mass of LBGs hosts inferred from clustering analysis, i.e. $M_h\sim3\times10^{11} - 10^{12}\rm~M_\odot$ (e.g. \citealt{2006ApJ...637..631K}). 
Since we expect the dependence of $\mathcal{T}_{IGM}$ on halo mass to extends in the range $11<\log_{10}M_h/{\rm M_\odot}<12$, we assume the sampling of the $\mathcal{T}_{IGM}-M_h$ relation at low mass haloes to mimic the realistic host halo mass of observed LBGs.

\label{lastpage}

\end{document}